\documentclass[sigconf,natbib=false]{acmart}

\acmConference[]{}{}{}
\usepackage{cite}
\usepackage{amsmath,amsfonts}
\usepackage{listings, color}
\usepackage{enumitem}
\usepackage{algorithmic}
\usepackage{graphicx}
\usepackage{textcomp}
\usepackage{xcolor}
\usepackage{tikz-cd}
\usepackage{tabularx}
\usepackage{listings, color}
\usepackage{multirow}
\usepackage[utf8]{inputenc}
\usepackage[english]{babel}
\usepackage[T1]{fontenc}
\usepackage[resetlabels,labeled]{multibib}
\usepackage{proof}
\usepackage{caption}
\usepackage{subcaption}

\setcopyright{none}
\renewcommand\footnotetextcopyrightpermission[1]{} % removes footnote with conference information in first column
\definecolor{codegreen}{rgb}{0,0.6,0}
\definecolor{codegray}{rgb}{0.5,0.5,0.5}
\definecolor{codepurple}{rgb}{0.58,0,0.82}
\definecolor{backcolour}{rgb}{0.95,0.95,0.92}

\def\BibTeX{{\rm B\kern-.05em{\sc i\kern-.025em b}\kern-.08em
    T\kern-.1667em\lower.7ex\hbox{E}\kern-.125emX}}

\newcommand{\forme}[1]{\textcolor{blue}{#1}}

\newif\ifanonymous
\anonymoustrue

\lstdefinestyle{PlainStyle}{
    basicstyle=\footnotesize,
    breakatwhitespace=true,         
    breaklines=true,                 
    showstringspaces=false,
    gobble=8,
    tabsize=2,
    frame=shadowbox,
    breakautoindent=false,
}

\lstdefinestyle{GrokStyle}{
    backgroundcolor=\color{backcolour},   
    commentstyle=\color{codegreen},
    keywordstyle=\color{magenta},
    numberstyle=\tiny\color{codegray},
    stringstyle=\color{codepurple},
    basicstyle=\footnotesize,
    breakatwhitespace=false,         
    breaklines=true,                 
    captionpos=b,                    
    keepspaces=true,                 
    numbers=left,                    
    numbersep=5pt,                  
    showspaces=false,                
    showstringspaces=false,
    showtabs=false,                  
    tabsize=2
}
    
\begin{document}

\title{Comprehending Variability in Analysis Results of Software Product Lines}
%\title{Question-Model Fitting - Literature review}

% \author{\IEEEauthorblockN{Rafael F. Toledo}
% \IEEEauthorblockA{\textit{University of Waterloo}}
% \and
% \author{\IEEEauthorblockN{Max}
% \IEEEauthorblockA{\textit{University of Waterloo}}
% \and
% \IEEEauthorblockN{Joanne M. Atlee}
% \IEEEauthorblockA{\textit{University of Waterloo}}
% }

\author{Rafael F. Toledo \hspace{1em} Joanne M. Atlee \hspace{1em} Rui Ming Xiong}
\email{{rftoledo, jmatlee, rmxiong}@uwaterloo.ca}
\affiliation{%
  \institution{University of Waterloo}
  \country{Canada}\\[2em]
}

\begin{abstract}
Analyses of a software product line (SPL) typically report variable results that are annotated with logical expressions indicating the set of product variants for which the results hold. These expressions can get complicated and difficult to reason about when the SPL has lots of features and product variants. Previous work introduced a visualizer that supports filters for highlighting the analysis results that apply to product variants of interest, but this work was weakly evaluated. In this paper, we report on a controlled user study that evaluates the effectiveness of this new visualizer in helping the user search variable results and compare the results of multiple variants. Our findings indicate that the use of the new visualizer significantly improves the correctness and efficiency of the user’s work and reduces the user’s cognitive load in working with variable results. 

%Results of variability-aware analyses of software product lines can be cumbersome due to their size and logical complexity. Graphical software models can ease comprehension by encoding connectivity aspects of the code elements and software variability. This work investigates how software models with edge group visualization can facilitate engineers' experience in comprehending variability-aware analysis results. Our results show that colour-marking the links between model elements reduces the time and mental effort required to comprehend analysis results, especially for comparing the analysis results of two or more variants. 
\end{abstract}
\maketitle

% \begin{IEEEkeywords}
% Software product lines, Variability-aware analyses, Analysis visualizations
% \end{IEEEkeywords}

\section{Introduction}
Developers pose software queries 
%(especially about control-flow, dataflow, and dependencies) 
in their efforts to understand the behaviour of large, complex software systems \cite{latoza2010developers,sillito2008asking}. It is well known that such queries can be difficult and time-consuming, and several researchers have investigated tools to ease both the asking of such questions and the understanding of the query results~\cite{barnett2014get, latoza2011, yoo2022investigating}.

These challenges become even more difficult when the software in question is a \textit{software product line (SPL)}. An SPL represents a \textit{family} of related software product \textit{variants} 
%(e.g., different models of cellphones or automobiles sold by the same company) 
and is typically structured as a collection of mandatory and optional features~\cite{Clements:2001}; product variants are distinguished by the set of optional \textit{features} they include. Thus, applying software queries to an SPL produces \textit{variable} results: many query results will apply to some product variants and not to others (e.g., if the query refers to feature-specific code). 

The typical approach to representing variable analysis results is to annotate each result with a \textit{presence condition (PC)}, where the PC represents the product variants 
%(in terms of a Boolean expression over feature variables, specifying the included and excluded features) 
for which the result holds~\cite{spllift-PC,ModelChecking-PC,TypeSafety-PC,Analysis-PC,ModelChecking-PC2,shahin2023applying}. A PC is a propositional formula over Boolean \textit{feature variables}, whose values denote whether their respective features are present ($true$) or not ($false$) in the PC's variants. 
%A PC can represent a single variant or
%by expressing the value of every feature variable or can represent 
%a set of variants. 
%For example, the PC $F$ represents all variants that include the feature F, and the PC $true$ represents all the SPL’s variants. 
Reasoning about variable analysis results is more demanding than simply matching PC expressions. If an engineer is interested in results pertaining to particular variants of concern (e.g., variants $(F\wedge !G)\!\vee\!H)$), the engineer needs to ascertain which of the many results have PC annotations that overlap with the variants of concern (i.e., which conjunctions of PCs and the variants-of-concern expression are \textit{satisfiable}).

Recently, Shahin et al.~\cite{shahin2023applying} presented enhancements to the Neo4j Browser~\cite{Neo4j} to not only annotate variable query results with their respective PCs but also provide support for highlighting results whose PCs overlap with a provided filter (also specified as a PC). The engineer can create multiple filters, assign each filter a different colour, and the browser will highlight each query result whose PC has at least one variant in common with the filters.
%and the filter expression have at least one variant in common. 
The enhanced browser looks promising in helping engineers to search and explore variable analysis results and compare the results of multiple variants, but the work has only been semi-formally evaluated by a small user study involving six engineers who commented on the perceived utility of images from the browser~\cite{shahin2023applying}.

We obtained the enhanced Neo4j Browser~\cite{shahin2023applying} from its authors and conducted a controlled user experiment to assess its effectiveness helping users reasoning about variable results from analyses of a software product line. Our study involved 42 student participants and found that the participants who used filters to highlight variant-specific results exhibited statistically significant improvements in the correctness and efficiency of their work and reported significantly lower levels of expended mental effort. A high majority of all participants reported a strong preference for such coloured filters, but lower-level visualization options of how variable results are coloured had no significant impact on performance. Although we evaluated only the enhanced Neo4j Browser, we expect that the visualizations and our study's findings would be applicable to other graph-based analyses and visualizers~\cite{latoza2011,yoo2022investigating} when applied to SPLs.

The rest of this paper is organized as follows. Section~\ref{sec:design} provides an overview of software product lines, analysis of SPLs, and the visualizer~\cite{shahin2023applying} being evaluated in this paper. Section~\ref{sec:eval} describes the design of our user study and we report the study results in Section~\ref{sec:results}. We conclude with a discussion of our take aways from the study and thoughts on potential directions for future work.

\section{Preliminaries}
\label{sec:design}

In this section, we overview fundamental concepts of software product lines (SPLs), analyses of SPLs, and the enhanced Neo4j Browser~\cite{shahin2023applying} used to visualize the variable results of SPL analyses.

% justify clearly the visualization
% - how to make the case that the structure you are visually showing benefits the target end user
% \textbf{TODO: }Present the data transformation from the code -> graph model -> analysis results -> graph with filter

%The interface presented in~\cite{shahin2023applying} is designed to support program comprehension of \textit{variability-aware analysis} results. The interface is a modified version of Neo4j Browser, the open-source visualizer developed for the Neo4j graph database. The database stores the code facts extracted from software product lines written in C++. The original version of the Browser offers basic functionalities for querying the database and visualizing the results. To support the visualization of variability-aware analysis results, Shahin et al.~\cite{shahin2023applying} augmented the Browser to enable the highlighting of results belonging to system variants specified by the user.

\subsection{Software Product Lines}

A \emph{software product line (SPL)} is a family of related software products or \textit{variants} that is typically structured as a collection of mandatory and optional features (e.g., cellphones, automotive software)~\cite{Clements:2001}. 
%The unit of variability in an SPL is a \emph{feature}, and 
An SPL's variants are differentiated by their optional features. The number of an SPL's variants can be exponential in the number of its optional features, but constraints among features typically preclude all possible feature combinations from being valid variants. 

\begin{figure}[t]
\includegraphics[width=0.48\textwidth]{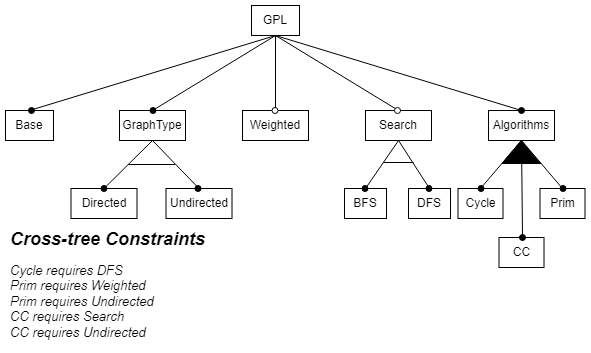}
%\vspace{-0.1in}
\caption{Feature model of a reduced version of GPL~\cite{lopez2001standard}.}
\label{fig:featureModel}	
\vspace*{-1em}
\end{figure}

Consider a graph product line (GPL)
%\footnote{\forme{https://github.com/paperMaterial4/icse2024}} 
(based on Lopez-Herrejon et al.~\cite{lopez2001standard}). The product line
%system is a family of graph applications that 
includes features implementing types of graphs (i.e., \textit{Directed}, \textit{Undirected}, \textit{Weighted}), search algorithms (i.e., \textit{breadth-first search (BFS)} and \textit{depth-first search (DFS)}), and other common graph algorithms (\textit{Cycle checking}, \textit{Connected Components (CC)}, and \textit{Prim's algorithm for minimum spanning tree}). 
%The GPL is designed to use graph concepts that are familiar to computer scientists, yet represent a non-trivial SPL %enough to explore fundamental concepts of software product lines, (e.g., not all features are compatible). 
A feature model~\cite{Kang:1990} (in Figure~\ref{fig:featureModel}) shows the GPL's features and feature constraints.
%Our version of a GPL\footnote{\forme{https://github.com/toledorafael/graphProductLine}} includes features implementing types of graphs (i.e., \textit{Directed}, \textit{Undirected}, \textit{Weighted}), search algorithms (i.e., \textit{breadth-first search (BFS)} and \textit{depth-first search (DFS)}), and other common graph algorithms (\textit{Cycle Checking}, \textit{Connected Components}, and \textit{Prim's algorithm for Minimum Spanning Tree (MST)}). Non-implemented features from the original specification of GPL~\cite{lopez2001standard} have constraints that are similar in complexity to those we chose to implement. For example, the \textit{Single-Source Shortest Path} algorithm, which is not included in our version of GPL, requires undirected weighted graphs, whereas \textit{Prim's MST} works only on directed weighted graphs. The configurable program and its variants are specified by the feature model in Figure~\ref{fig:featureModel}. 
The root node represents the full GPL and other nodes represent mandatory (black dot) and optional (white dot) features. The arcs underneath a node represent feature constraints between the node and its child nodes: if the node is in a variant, then a black arc indicates \textit{at least one} child node must also be included in the variant and a white arc indicates \textit{exactly one} child node must be included in the variant. Thus a GPL variant must include at least one graph algorithm and exactly one search algorithm.
%isthe child nodes are alternative subfeatures (at least one  features Cycle, CC, and Prim, indicates that at least one of Algorithms' child features must be included in a generated program variant. The white arc between the features breadth-first search (BFS) and depth-first search (DFS) indicate that only one of them can be enabled in a software variant.
%Features can be mandatory (black dot) or optional (white dot) in a program variant, and child nodes indicate the variations of the feature represented by the parent node. The legend in the bottom-right corner shows the symbols used to define constraints on the configurability of a parent node in terms of child-node selections. The black arc between the features Cycle, CC, and Prim, indicates that at least one of Algorithms' child features must be included in a generated program variant. The white arc between the features breadth-first search (BFS) and depth-first search (DFS) indicate that only one of them can be enabled in a software variant. 
The feature model also includes textual cross-tree constraints (shown in the bottom-left corner) that define constraints among features that are not directly related in the model's tree structure. A ``requires" constraint expresses a feature dependency and an ``excludes" constraint expresses mutual exclusion.
%In this configurable program, for example, variants that include the feature Cycle must also include DFS as its search algorithm.

\begin{figure*}[]
    \centering
    \begin{subfigure}[]{0.45\textwidth}%
	\lstset{style=GrokStyle}
    \begin{lstlisting}[language=C++,numbers=left]
void GraphApp::BFS(int srcID) {
    ...
    if (Weighted) {
        ...
        targetID = edge->getTargetID();
        
        if (Undirected) {
            ...
            srcID = edge->getSrcID();
        }
    }
    
    if (!Weighted){
        for (int neighborID : 
            nodes[currentNodeID]->getNgbrs()) {
            ...
        }
    }
...
    \end{lstlisting}
%    \vspace{-0.1in}
	\caption{Function \texttt{BFS}.}
	\label{fig:splCode}	
    \end{subfigure}
    \hspace*{2em}
    \begin{subfigure}[]{0.45\textwidth}%
	\includegraphics[width=\textwidth]{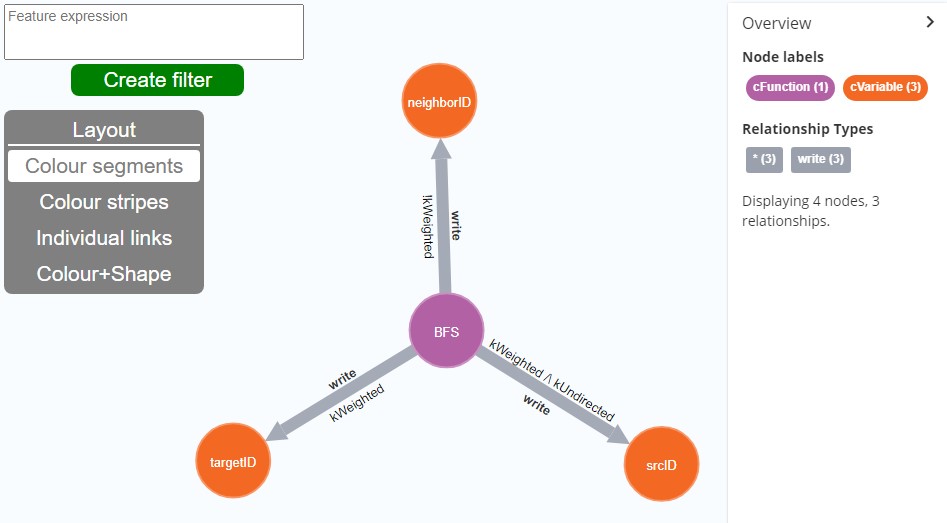}
%	\vspace{-0.1in}
	\caption{Dataflows in function \texttt{BFS}.}
 % MATCH (a:cFunction)-[b:write]->(c) WHERE a.label CONTAINS "BFS" AND NOT c.label CONTAINS "currNodeID" AND NOT c.label CONTAINS "queue" AND NOT c.label CONTAINS "edge" RETURN *
	%\vspace{-0.2in}
	\label{fig:splModel}
    \end{subfigure}
%\vspace
\caption{Code and dataflows annotated with PCs}
\label{fig:splExample}
\end{figure*}

In an \emph{annotative} SPL~\cite{Thum:2014}, feature-specific code is encapsulated in conditional statements that are guarded (annotated) with feature expressions. Each feature is represented by a Boolean constant (called a \emph{feature variable}), and \emph{feature expressions} are Boolean expressions over the feature variables. For example, Figure~\ref{fig:splExample}a shows snippets of the GPL's configurable BFS algorithm. The values of feature variables $Weighted$ and $Undirected$ (\emph{true} or \emph{false}) indicate whether their corresponding features are present or absent in a GPL variant. Lines 4-10 are specific to variants in which the feature $Weighted$ is present; lines 8-9 are specific to variants in which features $Weighted$ and $Undirected$ are both present; and lines 14-17 are specific only to variants in which feature $Weighted$ is absent.

%A specific value assignment to all of the SPL's feature variables is called a \emph{(feature) configuration} and denotes a single variant in the SPL. 
Variants can be represented by \emph{presence conditions (PCs)}, where a PC is a propositional formula over the feature variables\footnote{We use the syntax $!$, $\wedge$, $\vee$ for the propositional operators $not$, $and$, $or$, respectively.}. 
A PC can represent a single variant or
%by expressing the value of every feature variable or can represent 
a set of variants. 
Thus in the GPL, the PC \texttt{Weighted} represents all variants that include weighted graphs, and the PC $true$ represents all the GPL’s variants. 

\subsection{SPL Analysis and Visualization}

% What happens when analyzing SPL 
%Program analyses applied to an SPL analyze program data that may not be present in every SPL variant. 
Several program analyses have been \textit{lifted} to enable efficient analyses of SPLs~\cite{spllift-PC,ModelChecking-PC,TypeSafety-PC,Analysis-PC,ModelChecking-PC2,shahin2023applying}. Such analyses are called \textit{variability-aware analyses}, and they report results annotated with PCs representing the variants for which the results hold.
%are expected to preserve the semantics of single-product analysis while tracing each of the results to the set of products to which it applies. 
For example, the variability-aware analysis employed in Shahin et al.~\cite{shahin2023applying} is a lifted Datalog engine. Because the Datalog engine is lifted, rather than a specific query, this work supports any analysis that can be encoded as a Datalog query, including reachability analyses about control- and data-flows~\cite{shahin2023applying}, pointer and taint analyses~\cite{ShahinTSE}, and def-use analysis~\cite{datalog-defuse}. To enable analysis of large, heterogenous SPLs, this approach analyzes a \textit{factbase} model comprising a collection of \textit{program facts} extracted from the SPL's source code. Program facts include entities (e.g., functions, variables, components, control blocks), relationships between entities (e.g., function calls, variable assignments, containment, consecutive control blocks), and attributes of entities and relationships (e.g., names, PCs). Facts can be extracted from multiple types of artifacts written in different languages~\cite{davis2010whence,CPPX,bravenboer2009strictly,muscedere2019detecting}, and factbase analyses typically scale better than code analyses. 
%runs declarative analyses over models of annotative SPLs; that is, the annotations on the model edges representing the conditions guarding C/C++ code relations (e.g., function calls, variable writes) are processed and included on the analysis results as their presence conditions.

Consider again the code snippets in 
%consider an analysis detecting the potential data flows resulting from the execution of a specific function. A function \texttt{X} writes to a variable \texttt{a} if the function executes an assignment to that variable, updating its value. 
Figure~\ref{fig:splCode}.
%shows the body of function \texttt{BFS}, which includes variable assignments belonging to different GPL variants: 
The variable assignment to \texttt{targetID} (line 5) is in variants $Weighted$, the assignment to \texttt{srcID} (line 9) in in variants $Weighted\wedge\!Undirected$, and the assignment to \texttt{neighborID} (line 14) is in variants $!Weighted$. %But these writing relationships exist in different variants of the SPL. The write to \texttt{targetID} exists on variants with the feature \texttt{Weighted}, the write to \texttt{srcID} exists on variants including features \texttt{Weighted} and \texttt{Undirected}, and \texttt{neighborID} is updated on each iteration of a for-loop executed by variants without the feature \texttt{Weighted}. 
Figure~\ref{fig:splModel} shows an annotated graphical model representing these program facts and their presence conditions.
% Example
If an engineer wants to find dataflows in weighted, undirected variants of the GPL, they need to check that a dataflow's PC is satisfiable when the features $Weighted$ and $Undirected$ are enabled. %Identifying the write edges with satisfiable presence conditions determines the variable assignments executing in that group of variants. 
In this example, the \texttt{write} facts involving \texttt{targetID} and \texttt{srcID} hold in the variants of interest.

\label{sec:customization}

\begin{figure}[]
	\centering
	\includegraphics[width=0.48\textwidth]{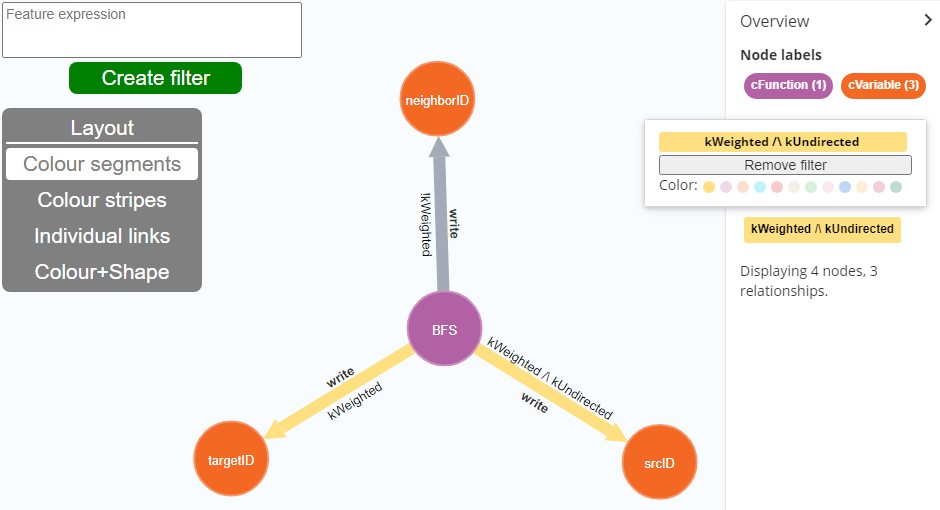}
	%\vspace{-0.1in}
	\caption{Highlighted \texttt{BFS} dataflows in variants of interest}
	%\vspace{-0.2in}
	\label{fig:filteredInterface}
\end{figure}

Shahin et al.~\cite{shahin2023applying} introduces a graph-based interface that not only visualizes PC-annotated results of variability-aware factbase analyses, but also provides facilities to highlight results that are relevant to the engineer. 
%Figure~\ref{fig:splModel} shows the components of their interface. 
The user types the PCs of the variants of interest into a textbox in the top-left corner of the interface and clicks the green button to create a filter. The interface then colours the model elements whose PCs have at least one variant in common with the filter's PC. 
%The engineer, considered in the previous example, looking for data flows in GPL variants using weighted undirected graphs, could create a filter with feature configuration \texttt{kWeighted} $\vee$ \texttt{kUndirected} to identify relevant write relations. 
Figure~\ref{fig:filteredInterface} shows the result of applying a filter to the dataflows visualized in Figure~\ref{fig:splModel}. 
%Only write relations between \texttt{BFS} and variables \texttt{targetID} and \texttt{srcID} are coloured in yellow. 

\begin{figure*}[t]

\begin{subfigure}[]{0.24\textwidth}
	\centering
	\includegraphics[width=\textwidth]{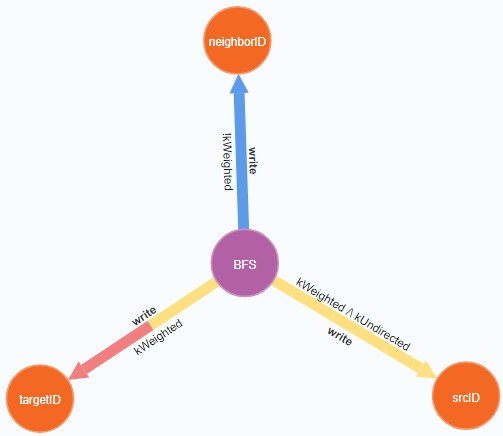}
	%\vspace{-0.1in}
	\caption{Arrow segments}
	%\vspace{-0.2in}
	\label{fig:segments}
\end{subfigure}
% \hfill
\begin{subfigure}[]{0.24\textwidth}
	\centering
	\includegraphics[width=\textwidth]{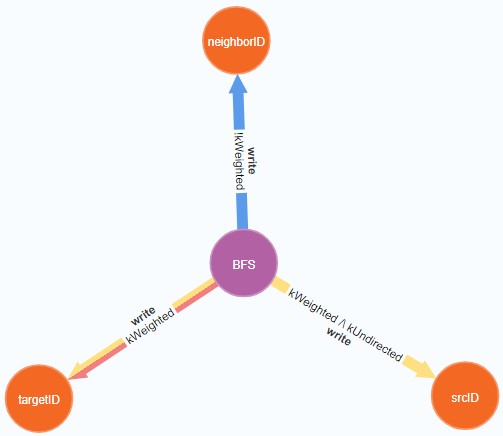}
	%\vspace{-0.1in}
	\caption{Arrow stripes}
	%\vspace{-0.2in}
	\label{fig:stripes}
\end{subfigure}
% \hfill
\begin{subfigure}[]{0.24\textwidth}
	\centering
	\includegraphics[width=\textwidth]{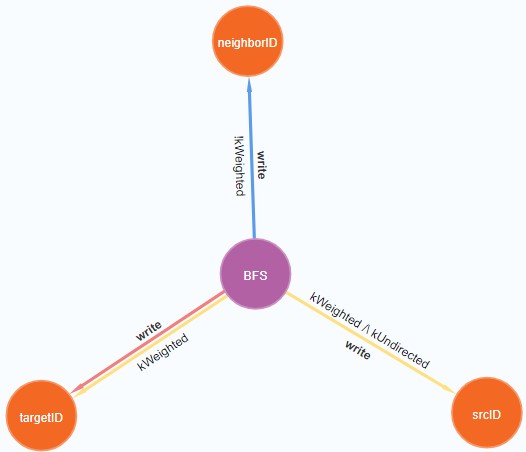}
	%\vspace{-0.1in}
	\caption{Separate arrows}
	%\vspace{-0.2in}
	\label{fig:separateArrows}
\end{subfigure}
%\hfill
\begin{subfigure}[]{0.24\textwidth}
	\centering
	\includegraphics[width=\textwidth]{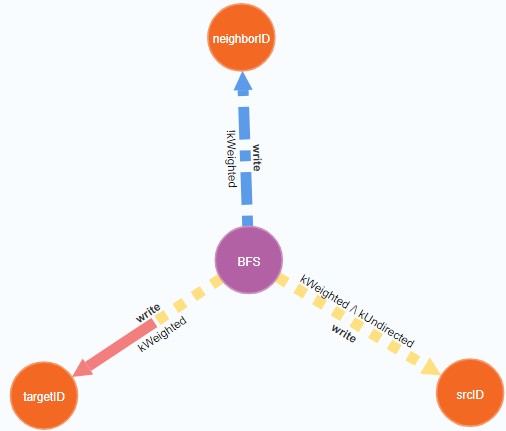}
	%\vspace{-0.1in}
	\caption{Arrows shapes}
	%\vspace{-0.2in}
	\label{fig:shapes}
\end{subfigure}
\caption{Highlight visualization options}
\label{fig:filtersExample}
%\vspace{-0.2in}
\end{figure*}

The engineer can create multiple filters to compare variants and can customize the colour of each filter's highlights.
%using a menu on the right side of the interface and can customize how highlighted results are visualized via a menu on the left. 
They can also customize how highlights from multiple filters are visualized: (a)~\textit{Arrow segments}, (b)~\textit{Arrow stripes}, (c)~\textit{Separate Arrows}, and (d)~\textit{Arrow shapes}. Figures~\ref{fig:filtersExample}a-d show examples of these four options.
%A gray box on the left side of the model, as shown in Figure~\ref{fig:filteredInterface}, lists all filter visualization that the user may select. 

The interface is a promising approach for visualizing the results of variability-aware analyses but its usability has been weakly evaluated.  Shahin et al.~\cite{shahin2023applying} conducted an unobserved experiment with six engineers that assessed the engineers' impressions of images from the interface. To address this weakness, we conducted an observed experiment to measure and assess the experience of many more participants who used the visualizer's features to understand and reason about variability-aware analysis results.

%The visualization option changes how the filters' colours are distributed and represented in the model edges. \textit{Arrow Segments} distribute the colour marks through the extension of the arrow from the base to the head. \textit{Arrow Stripes} distribute the colour marks from side to side of the arrow. \textit{Separate Arrows} replace the analysis results with a set of smaller arrows, each representing a filter that satisfies that relation's PC. \textit{Arrow Shapes} apply the distribution of colours similarly to the arrow segments, and it also allows the customization of the shape of each arrow segment. Figure~\ref{fig:filtersExample} shows the variations in filter visualization provided by the interface.

\section{Experimental Design}
\label{sec:eval}

%Previous work has introduced a promising but weakly evaluated interface for visualizing the results of variability-aware analyses~\cite{shahin2023applying}. Their evaluation consists of an unobserved experiment with few participants and focuses on the participants' impressions of images from the interface. In contrast, this work is a more complete evaluation involving an observed experiment measuring and assessing the experience of many more participants who used the visualizer's features to understand and reason about the analysis results.

Our evaluation of the enhanced Neo4j Browser considers multiple aspects of the user's experience. Our primary goal was to assess hypotheses about gains in time, correctness, and cognitive load that the visualization features might provide and whether the filter customization options would affect any measured gains. We also wanted to understand how the users feel about the interface.

We obtained a copy of the interface from the authors of \cite{shahin2023applying} and designed a controlled user study that compares the original Neo4j Browser~\cite{Neo4jBrowser} and the variability-aware visualizer presented in \cite{shahin2023applying} with respect to the following null hypotheses. 

% Transition from Introduction to hypotheses about benefits of UI
\begin{enumerate}[label=H\arabic*]
    % Significance level 1%
    % (metric: task time)
    \item \textbf{Efficiency:} Using coloured filters does not change the amount of time it takes to comprehend variability-aware analysis results. 
    % (metric: error rate)
    \item \textbf{Correctness:} Using coloured filters does not change the number of mistakes made when comprehending variability-aware analysis results.
    % (metric: [the SMEQ scale](https://www.bentley.edu/centers/user-experience-center/measuring-difficulty-doesnt-need-be-difficult))
    \item \textbf{Cognitive load:} Using coloured filters does not change the cognitive load of performing comprehension tasks involving variability-aware analysis results.
    % (metric: Likert scale + 100-dollars preference)
    \item \textbf{User preference:} Users feel indifferent about using coloured filters to highlight variability-aware analysis results.  
    % (compare all metrics between visualization options)
    \item \textbf{Filter Visualization Options:} The filters visualization options have the same impact on user efficiency and correctness.

\end{enumerate}

Study sessions were held online during one-on-one video calls between each participant and the researcher. Participants accessed the study materials (tools, tasks, survey) via a web browser on their computer (i.e., laptop or desktop), and they shared their screen with the researcher. 

\begin{figure}[t]
\includegraphics[width=0.34\textwidth]{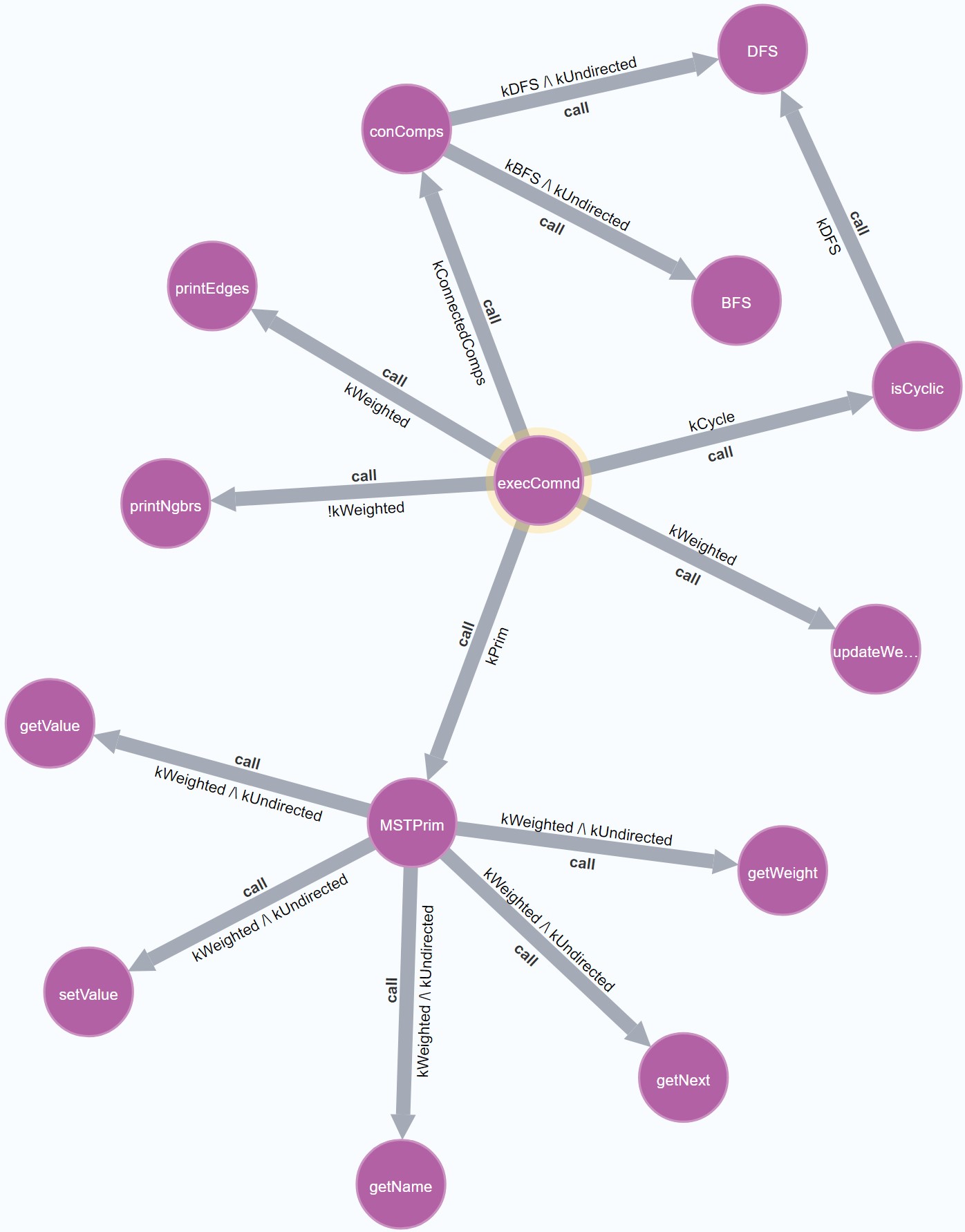}
% \vspace{-0.1in}
\caption{Graphical model of the results from a variability-aware control-flow analysis of GPL function \texttt{execComnd}.}
\label{fig:taskModel}	
\end{figure}

\begin{figure}[]
    % \centering
    \begin{subfigure}[]{0.48\textwidth}
        \lstset{style=PlainStyle}
        \begin{lstlisting}
        Which function(s) may be directly called by the function execComnd in the following program variant V1?
        V1: kWeighted /\ kUndirected /\ kDFS /\ !kBFS /\ kCycle /\ kConnectedComps /\ !kPrim
        \end{lstlisting}
        \vspace{-0.05in}
        \caption{Task example for identifying facts belonging to a single variant.}
        \label{fig:identifyExample}
        \vspace*{1em}
    \end{subfigure}
    \begin{subfigure}[]{0.48\textwidth}
        \lstset{style=PlainStyle}
        \begin{lstlisting}
        Which function(s) may be directly called by the function execComnd in the program variant V2 but not in V1?
        V1: kWeighted /\ kUndirected /\ kDFS /\ !kBFS /\ kCycle /\ kConnectedComps /\ !kPrim
        V2: kWeighted /\ kUndirected /\ !kDFS /\ kBFS /\ !kCycle /\ kConnectedComps /\ kPrim
        \end{lstlisting}
        \vspace{-0.05in}
        \caption{Task example for comparing facts from two different variants.}
        \label{fig:compare2Example}
    \end{subfigure}

    % \begin{subfigure}[]{0.48\textwidth}
    %     \lstset{style=PlainStyle}
    %     \begin{lstlisting}
    %     There are two possible call paths between the functions execComnd and DFS. Which of program variants may execute both call paths?
    %     V1: kWeighted /\ kUndirected /\ kDFS /\ !kBFS /\ kCycle /\ kConnectedComps /\ !kPrim
    %     V2: kWeighted /\ kUndirected /\ !kDFS /\ kBFS /\ !kCycle /\ kConnectedComps /\ kPrim
    %     V3: kWeighted /\ kUndirected /\ !kDFS /\ kBFS /\ !kCycle /\ !kConnectedComps /\ kPrim
    %     \end{lstlisting}
    %     \vspace{-0.1in}
    %     \caption{Task example for comparing facts from three different variants.}
    %     \label{fig:compare3Example}
    % \end{subfigure}
    %\vspace{-0.2in}

    \caption{Examples of study task.}
    \label{fig:taskExample}
\end{figure}

\subsection{Task Design}

The study's tasks fall into three categories: identify facts belonging to a single product variant, compare facts from two different variants, and compare facts from three different variants. These tasks are applied to graphical models of a variability-aware analysis. Figure~\ref{fig:taskModel} shows a graphical model of the results from a variability-aware control-flow analysis starting from the GPL function \texttt{execComnd}.

Each study question provides a graphical model of variability-aware analysis results including PC annotations, the task to be performed, and the variants of interest for that task. Thus, a task to \textit{identify} facts in a single variant asks the participant to find instances of function calls or variable assignments that exist in the given variant. Such tasks simulate how an engineer might investigate control- and data-flows among functions or variables. Figure~\ref{fig:identifyExample} shows an example of this kind of task. A task to \textit{compare} the facts \textit{between} two different variants specifies two variants of interest and asks the subject to find specific facts in the intersection, union, or set-difference of the two variants (Figure~\ref{fig:compare2Example}). The set-difference of variants could represent the enabling or disabling of a single feature, simulating how an engineer might assess the impact of modifying a feature configuration. Lastly, a task to \textit{compare among} multiple variants extends a comparison task to three variants of interest.

The tasks of the study were divided into two stages. In the first stage, the participants interacted freely with a Neo4j database and visualizer via a publicly accessible webpage while sharing their screens with the researcher. Both the control group and the treatment group performed three tasks on each of the two models. Both groups had to run Neo4j queries (provided by the study) to create the tasks' models, and both groups used their respective visualizers to customize the models. Treatment participants were able to create coloured filters that highlight a variant's facts, whereas control participants could use only Neo4j's basic customization options.
%%%%%%%%%%%THESIS%%%%%%%%%%%%%%%%%
% We prohibited the participants from modifying the models they worked with during the study's second stage, to ensure that the treatment participants used all four visualization options, so that we could  test hypotheses $H4$ and $H5$.
%%%%%%%%%%%THESIS%%%%%%%%%%%%%%%%%

In the second stage of the study, the participants answered an online survey comprising twelve comprehension questions about four models. The models were screenshots  that the participants could not modify; and the four models provided to the treatment group covered each of the visualization options (i.e., $Arrow Stripes$, $Arrow Segments$, $Arrow Shapes$, and $Separate Arrows$). Different treatment participants experienced the visualization options in different orders, according to a balanced Latin square design~\cite{grant1948latin}, to decrease any effect the presentation order could have on the participants' performance. The control group performed identical tasks on versions of the models with uncolored edges. 

% The order in which they are presented to the participants varies according to the balanced Latin square to decrease any effect the presentation order could have on the study results. Each participant is randomly assigned to follow one of the possible sequences of edge-group visualization options:
% \begin{enumerate}
%     \item arrow stripes, arrow segments, arrow shapes, separate arrows
%     \item arrow segments, separate arrows, arrow stripes, arrow shapes
%     \item separate arrows, arrow shapes, arrow segments, arrow stripes
%     \item arrow shapes, arrow stripes, separate arrows, arrow segments
% \end{enumerate}

During the study, we discovered that one of the second-stage questions was phrased ambiguously, which confused some of the participants. Thus data collected from that particular question were removed from our analyses of the participants' performance.

The complete list of study questions and other materials like the screening questionnaire, pre-study tutorial, the GPL code base, the post-study feedback form, and study results can be found in an anonymous git repository\footnote{\forme{https://github.com/paperMaterial4/icse2024comprehending}} created for this paper.

\subsection{Recruitment}

We advertised the study to students in our university's second- and fourth-year Computer Science (CS) and Software Engineering (SE) courses. Students in these courses are expected to be familiar with Boolean logic and configurable software (e.g., using C preprocessor directives) from previous courses on logic and software design. In addition, most students at that level will have completed 1-4 internships at tech companies. To further ensure participants were proficient in these concepts, all 63 students who expressed interest in participating in the study completed a screening questionnaire that assessed their understanding of the graph product line (GPL) and models annotated with presence conditions. Students who correctly answered at least six of the seven screening questions were deemed eligible to participate in the study. Forty-two students were invited to schedule a study session. 

Every study session started with a written tutorial
%\footnote{\forme{https://github.com/paperMaterial4/icse2024}} 
that introduced the participants to the Neo4j Browser (original or variability-aware) and its customization options. The tutorial included a few demo tasks asking questions about small models (with 3-4 nodes) and instructing participants to use the customization options. It took participants about an hour to complete both the tutorial and the user study. Study participants received \$50 in appreciation for their participation.

\subsection{Application Domain}

The study's comprehension tasks are performed on the results of variability-aware queries of the graph product line (GPL) introduced in Section~\ref{sec:design}. The GPL was designed to use graph concepts that are familiar to second-year CS and SE students (thereby mitigating against accidental complexities induced by the application domain), yet represent a non-trivial SPL (e.g., not all features are compatible).

\subsection{Treatment Allocation}
% threat of selection bias(?)
We distributed the students into the control and treatment groups by alternating the inclusion of new participants in each group based on the initial schedule of study sessions. Whenever a participant had to reschedule, their allocation persisted to the group to which they were originally assigned. Alternating allocation was also applied to the distribution of treatment participants to the four different orderings of the filter visualization options.

\subsection{Data Collection and Metrics}

In order to test our hypotheses, during the study we collected the participants' solutions and times on tasks. \textit{Correctness} of a participant's solution was measured by comparing them to objectively correct answers. A solution was assessed as either correct or not, with no notion of partial correctness. \textit{Time on task} measures the time a participant spent working on a task before submitting an answer. The mental effort applied to each category of tasks is self-reported by participants using an online version of Subjective Mental Effort Questionnaire (SMEQ) scale~\cite{sauro2009comparison}. The scale has nine labels, ranging from \textit{``Not at all hard to do''} to \textit{``Tremendously hard to do''}, which are distributed over numerical scores ranging from 0 to 150. 

% \begin{figure}[t]
% \centering
% \includegraphics[width=0.30\textwidth]{SMEQScalePaper.png}
% %\vspace{-0.1in}
% \caption{Original SMEQ scale~\cite{sauro2009comparison}.}
% \label{fig:SMEQScalePaper}	
% \end{figure}

\begin{figure}[t]
\includegraphics[width=0.45\textwidth]{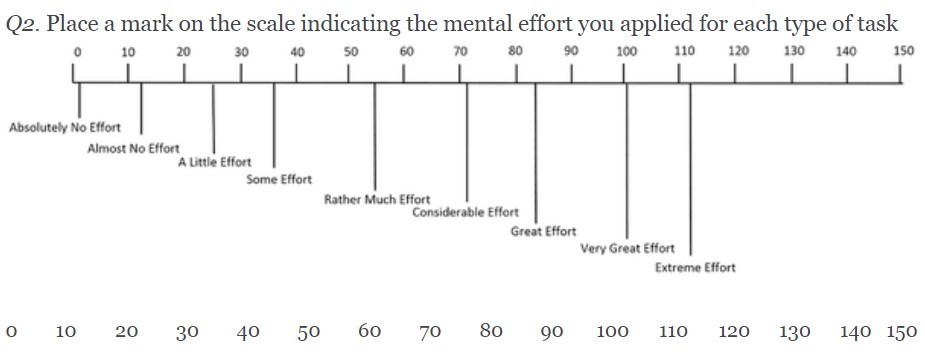}
%\vspace{-0.1in}
\caption{SMEQ scale used in the user study.}
\label{fig:SMEQScale}	
\end{figure}

Participants' self-reported mental effort and opinions of the visualizer were collected in a post-study feedback form. The form has four parts: 
\begin{enumerate}
    \item \textbf{Easy of Use:} Participants were given eight statements about the usability of the visualizer and asked to provide their level of agreement in terms of a Likert scale~\cite{allen2007likert}. Control and treatment received group-specific statements regarding their respective visualizers. For example, the first statement of the two feedback forms state "I found that interpreting the [\textit{Boolean expressions} vs. \textit{coloured filters}] were unnecessarily complex", respectively. The statements vary as to whether they are positive or negative about the user experience, to avoid influencing participants' answers.
    \item \textbf{Self-reported mental effort:} Participants reported the degree of mental effort they applied for each task category, using horizontal sliders ranging over the SMEQ scale~\cite{sauro2009comparison}, as shown in Figure~\ref{fig:SMEQScale}.
    \item \textbf{Ranked preferences for the visualization features:} Ranked preferences were collected using the \$100 prioritization method~\cite{leffingwell2000managing}. Control participants were asked to distribute a hypothetical \$100 dollars to missing features they would have liked to see in their visualizer tasks; options included the ability to customize Boolean expressions, to elide edges that do not apply in a variant, and to colour edges based on the variants to which they apply. Treatment-group participants did the same \$100 prioritization exercise, but they ranked the four visualization options for highlighting filter results.
    \item \textbf{Comments:} Two open-ended questions ask participants for comments on their experiences ("\textit{What are your general impressions of the presented tool?}) and for suggested improvements to the visualizer ("\textit{Do you have any suggestions for improvements to the presented tool? If so, please describe them below.}"). 
\end{enumerate}

\section{Evaluation Results}
\label{sec:results}
% 29 men; 12 women; 1 prefer to not say their gender

% 2 colourblind (both red-green, one of them claims partial colorblindness)

% Student Programs:

% Computer Science	12,
% Statistics	1,
% Material and Nanoscience,	1
% Computer Science and Business Administration Double Degree	1,
% Software Engineering	15,
% Computer Engineering	11,
% Mechatronics Engineering	1.

Of the 42 participants in our study, 38 are undergraduate students, and 4 are PhD students; 29 of the 42 are men. Two participants declared they have full or partial red-green colour blindness. Most participants (90\%) as majors in Software Engineering, Computer Science, or Computer Engineering; others are majors in Statistics, Mechatronics Engineering, or Nanoscience. All participants passed the screening questionnaire and had either completed or, at the time of the study, were enrolled in a course that taught configurable software methodologies (e.g., C preprocessor directives like \texttt{\#ifdef}, \texttt{\#ifndef}). Seventeen participants declared they have prior work experience with configurable software. None had previously used Neo4j. 

Three participants in the treatment group chose not to use the coloured filters in the first stage of the study. We will explore their reasons in the next sections, but given that this means they didn't experience the usage of the treatment visualizer, we removed their stage-one data from our analysis of participants' performance.

\subsection{Efficiency}

Our null hypothesis related to efficiency was:
\begin{enumerate}[label=H\arabic*]
\item\textbf{Efficiency:} Using the coloured filters does not change the amount of time it takes to comprehend program variability-aware analysis results. 
\end{enumerate}

\begin{figure}[t]
    % \begin{subfigure}[]{0.45\textwidth}
    % 	% \centering
    % 	\includegraphics[width=\textwidth]{charts/stage1Time.png}
    % 	%\vspace
    % 	\caption{Stage 1}
    % 	%\vspace{-0.2in}
    % 	\label{fig:stage1time}
    % \end{subfigure}

    % \begin{subfigure}[]{0.45\textwidth}
    %     % \centering
    %     \includegraphics[width=\textwidth]{charts/stage2Time.png}
    %     %\vspace{-0.1in}
    %     \caption{Stage 2}
    %     %\vspace{-0.2in}
    %     \label{fig:stage2time}
    % \end{subfigure}
    \includegraphics[width=0.45\textwidth]{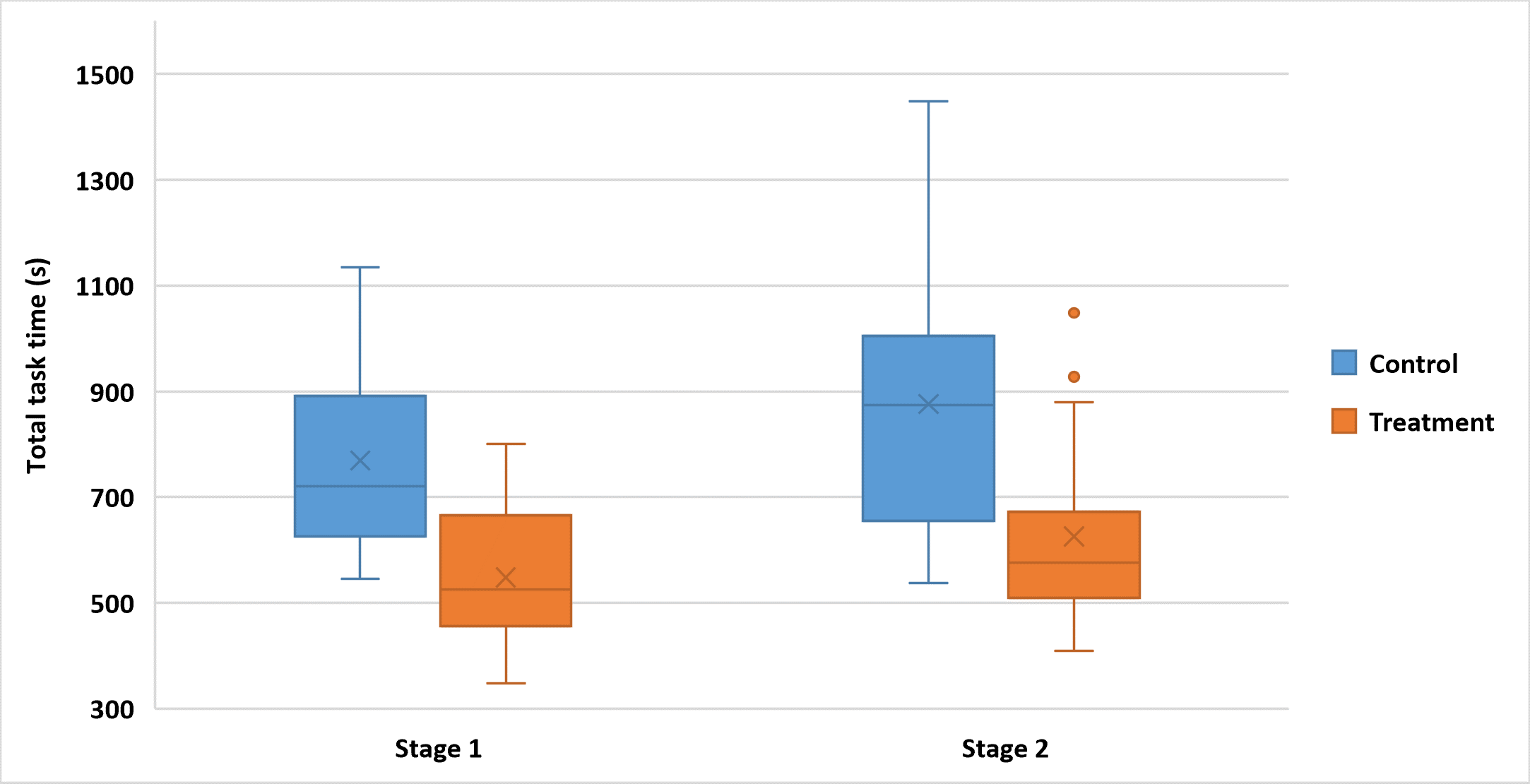}
%    \vspace{-0.1in}
    \caption{Average time to perform tasks, disaggregated by study stage.}
    \label{fig:stagesTime}
\end{figure}

% Both stages
\noindent This hypothesis is rejected ($p < 0.01$). The treatment participants finished all tasks approximately 30\% faster than the control participants. We observed similar time reductions in both stages of the study, suggesting that participants who used coloured filters took less time to perform study tasks regardless of whether they had the freedom to customize the model. Efficiencies observed in the first stage of the study also indicate that the time that treatment participants took to create and customize filters was offset by the reduced time it took to interpret the models. Figure~\ref{fig:stagesTime} shows summaries of the time the control and treatment groups took during the two study stages. The box-and-whisker plots on the charts show %measures of location and 
the distribution of results. The length of the box represents the interquartile range, the line through the box shows the median average, the \texttt{X} sign marks the mean average, and the whiskers indicate the maximum and minimum values observed for that sample. Data points exceeding $1.5$ times the interquartile range, are classified as outliers and are represented by dots above or below the boxes.   

\begin{figure}[t]
    % \begin{subfigure}[]{0.33\textwidth}
    % 	% \centering
    % 	\includegraphics[width=\textwidth]{charts/identifyTime.png}
    % 	%\vspace{-0.1in}
    % 	\caption{Identifying a single variant's results.}
    % 	%\vspace{-0.2in}
    % 	\label{fig:identifyTime}
    % \end{subfigure}
    % % \hfill
    % \begin{subfigure}[]{0.33\textwidth}
    %     % \centering
    %     \includegraphics[width=\textwidth]{charts/compare2Time.png}
    % 	%\vspace{-0.1in}
    % 	\caption{Comparing results of two variants.}
    % 	%\vspace{-0.2in}
    % 	\label{fig:compare2Time}
    % \end{subfigure}
    % % \hfill
    % \begin{subfigure}[]{0.33\textwidth}
    %     % \centering
    %     \includegraphics[width=\textwidth]{charts/compare3Time.png}
    % 	%\vspace{-0.1in}
    % 	\caption{Comparing results of three variants.}
    % 	%\vspace{-0.2in}
    % 	\label{fig:compare3Time}
    % \end{subfigure}
    \includegraphics[width=0.45\textwidth]{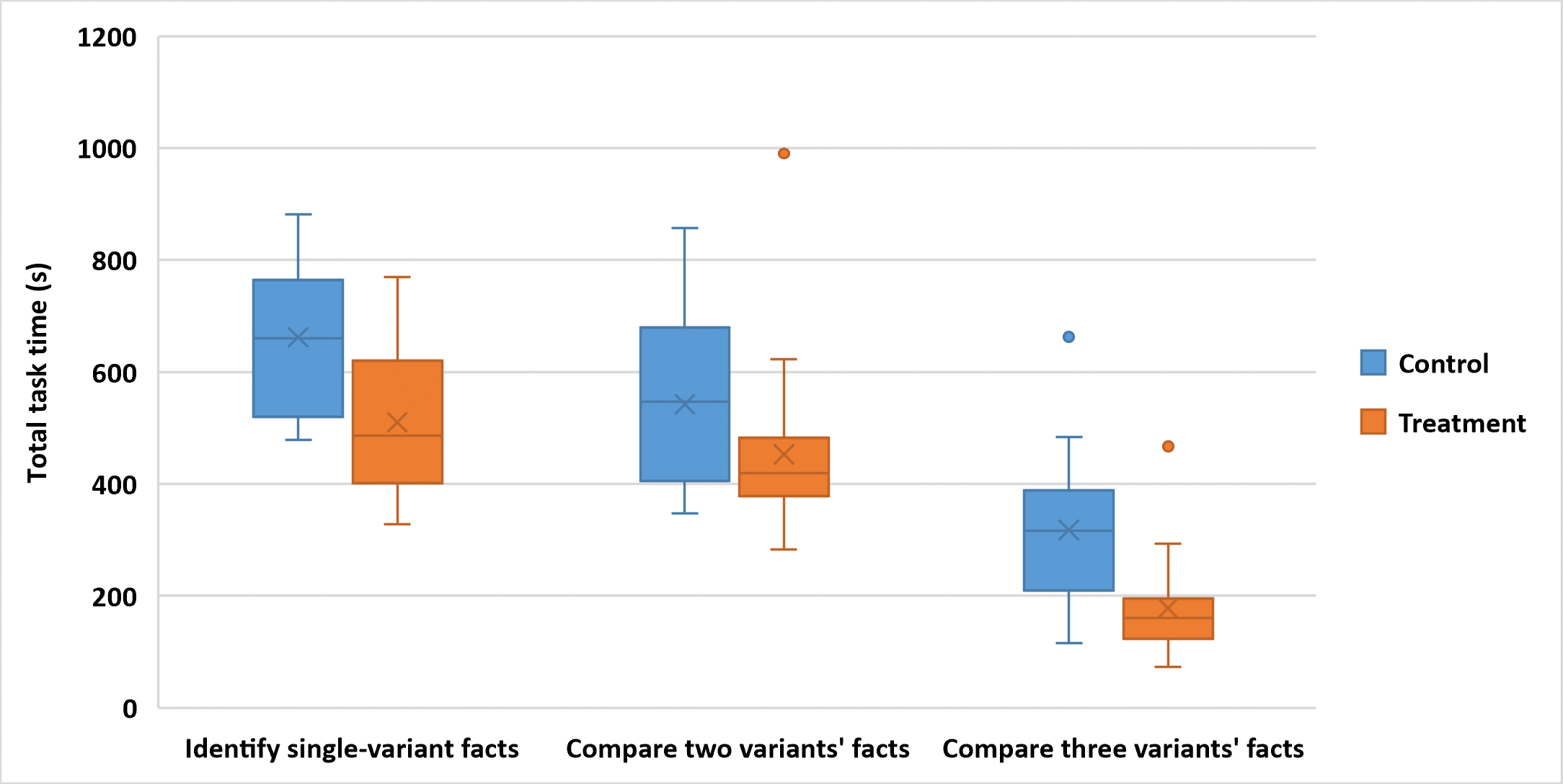}
%    \vspace{-0.1in}
    \caption{Average time to perform tasks, disaggregated by task category.}
    \label{fig:tasksTime}
\end{figure}

We see the same efficiency gains regardless in some of the type of task being performed. The treatment participants were faster in all task categories, but they were significantly faster when identifying facts belonging to a given variant and comparing facts from three variants ($p < 0.01$). Participants' average had a reduction of 22\% when working with a single variant and 44\% when multiple variants were considered. When comparing results from two variants, participants' average time was not significantly different from the control group ($0.05 < p < 0.1$). Figure~\ref{fig:tasksTime} compares the task times for both participant groups, disaggregated by the task category.

% \begin{figure}[t]
% 	\centering
% 	\includegraphics[width=0.48\textwidth]{charts/identifyTime.png}
% 	%\vspace{-0.1in}
% 	\caption{Time for tasks identifying a fact variant}
% 	%\vspace{-0.2in}
% 	\label{fig:identifyTime}
% \end{figure}

% \begin{figure}[t]
% 	\centering
% 	\includegraphics[width=0.48\textwidth]{charts/compare2Time.png}
% 	%\vspace{-0.1in}
% 	\caption{Time for tasks comparing facts of two variants}
% 	%\vspace{-0.2in}
% 	\label{fig:compare2Time}
% \end{figure}

% \begin{figure}[t]
% 	\centering
% 	\includegraphics[width=0.48\textwidth]{charts/compare3Time.png}
% 	%\vspace{-0.1in}
% 	\caption{Time for tasks comparing facts of three variants}
% 	%\vspace{-0.2in}
% 	\label{fig:compare3Time}
% \end{figure}

\subsection{Correctness}

Our null hypothesis related to correctness was:

\begin{enumerate}[label=H\arabic*]
\setcounter{enumi}{1}
\item\textbf{Correctness:} Using coloured filters does not change the number of mistakes made when comprehending variability-aware analysis results.
\end{enumerate}

\begin{figure}[t]
    % \begin{subfigure}[]{0.45\textwidth}
    % 	% \centering
    % 	\includegraphics[width=\textwidth]{charts/stage1Correctness.png}
    %     %\vspace{-0.1in}
    %     \caption{Stage 1}
    %     %\vspace{-0.2in}
    %     \label{fig:stage1Correctness}
    % \end{subfigure}

    % \begin{subfigure}[]{0.45\textwidth}
    %     % \centering
    %     \includegraphics[width=\textwidth]{charts/stage2Correctness.png}
    % 	%\vspace{-0.1in}
    % 	\caption{Stage 2}
    % 	%\vspace{-0.2in}
    % 	\label{fig:stage2Correctness}
    % \end{subfigure}
    \includegraphics[width=0.45\textwidth]{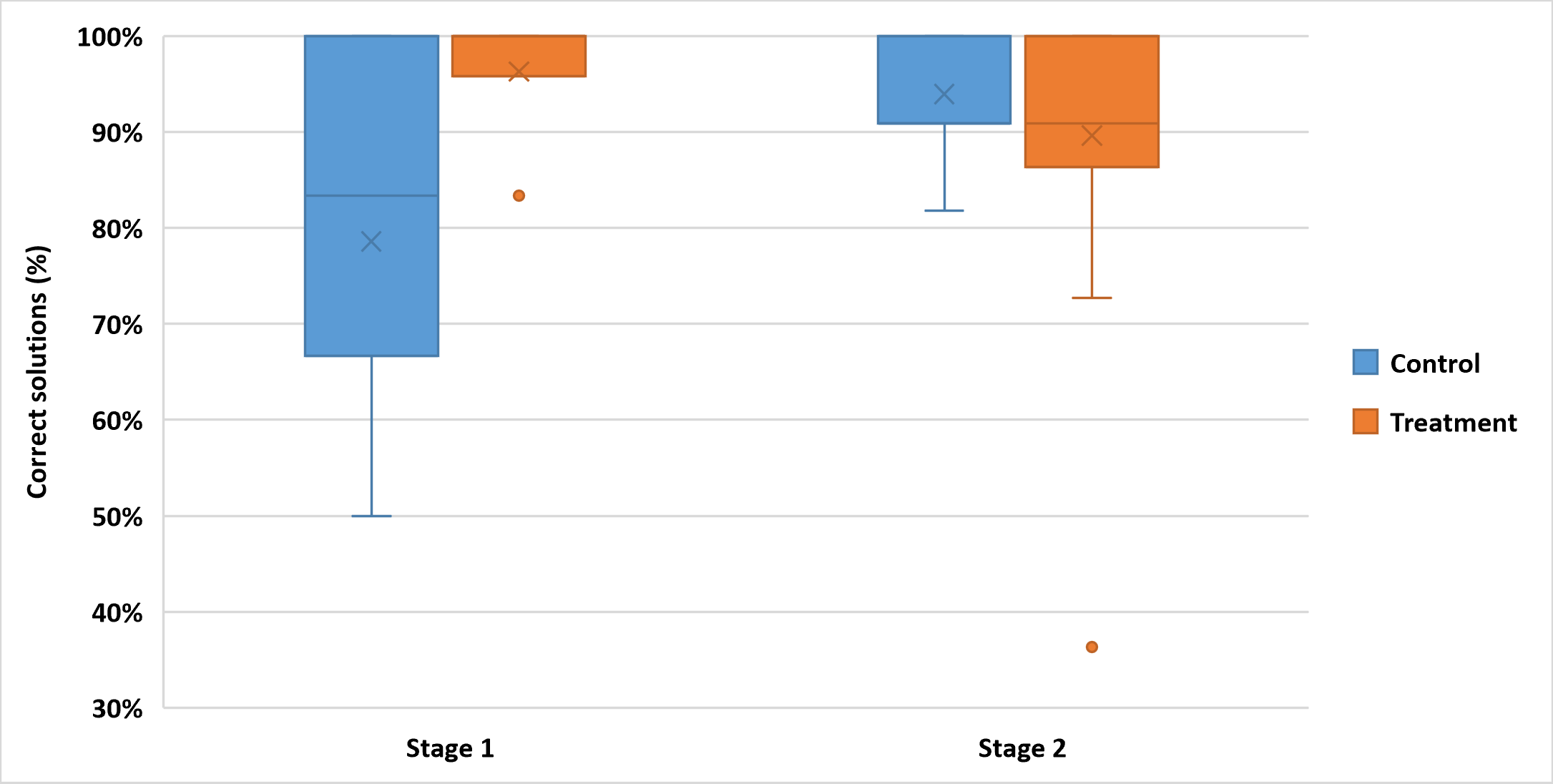}
%    \vspace{-0.1in}
    \caption{Average participant correctness, disaggregated by study stage.}
    \label{fig:stagesCorrectness}
\end{figure}

\noindent The hypothesis is rejected as the treatment participants made fewer mistakes, especially in the first stage of the study ($p < 0.01$), where participants from the control group had an average error rate almost six times higher than participants who used the coloured filters. Figure~\ref{fig:stagesCorrectness} shows the average accuracies of participants in both stages of the study. 

Looking in greater detail at the errors made, the most common error was failure to discern the direction of edges (or to notice that edges were directed), which is required to comprehend directed relationships (e.g., calling vs. called functions); 62\% of the control participants made such a mistake on at least one task, whereas 11\% of the treatment participants committed this same error. Control participants were also more likely to misinterpret the scope of a task when asked for singleton function calls made by a function of interest, a quarter of the control participants reported entire call chains starting from that function. None of the participants from the treatment group made such a mistake. Very few participants from either study group made mistakes in determining which model elements belong to the variants of interest.

In stage 2 of the study, the two groups' accuracy is not statistically significant ($0.2 < p$). Control participants performed better than in the first stage, and treatment participants performed slightly worse, making mistakes they did not make in the first stage; in particular, one outlier achieved only 36\% correctness in their stage 2 tasks. Recall that the differences between stage 1 and stage 2 tasks are twofold: (1) In stage one, participants interacted with the Neo4j tool, entering queries, possibly customizing the visualization of the query result, and answering variant-specific questions about the query results; whereas in stage 2, participants were given images of query results. (2) In stage two, the models provided to the control group all exhibit the same view of variable query results (PC annotation on model elements), whereas the models provided to the treatment group exhibited different options for highlighting variant-specific results. In Section~\ref{sec:discussion}, we discuss how these differences between the first and second stages of the study may have affected participants' performance. Despite the results from stage two of the study, we reject the null hypothesis based on the results from stage one, which was a more realistic assessment of participants' interactions with the study's visualizers. 

% \begin{figure}[t]
% 	\centering
% 	\includegraphics[width=0.48\textwidth]{charts/stage1Correctness.png}
% 	%\vspace{-0.1in}
% 	\caption{Stage 1 correctness}
% 	%\vspace{-0.2in}
% 	\label{fig:stage1Correctness}
% \end{figure}

% \begin{figure}[t]
% 	\centering
% 	\includegraphics[width=0.48\textwidth]{charts/stage2Correctness.png}
% 	%\vspace{-0.1in}
% 	\caption{Stage 2 correctness}
% 	%\vspace{-0.2in}
% 	\label{fig:stage2Correctness}
% \end{figure}

\begin{figure}[t]
    % \begin{subfigure}[]{0.33\textwidth}
    % 	% \centering
    % 	\includegraphics[width=\textwidth]{charts/identifyCorrectness.png}
    % 	%\vspace{-0.1in}
    % 	\caption{Identifying a single variant's results.}
    % 	%\vspace{-0.2in}
    % 	\label{fig:identifyCorrectness}
    % \end{subfigure}
    % % \hfill
    % \begin{subfigure}[]{0.33\textwidth}
    %     % \centering
    %     \includegraphics[width=\textwidth]{charts/compare2Correctness.png}
    % 	%\vspace{-0.1in}
    % 	\caption{Comparing results of two variants.}
    % 	%\vspace{-0.2in}
    % 	\label{fig:compare2Correctness}
    % \end{subfigure}
    % % \hfill
    % \begin{subfigure}[]{0.33\textwidth}
    %     % \centering
    %     \includegraphics[width=\textwidth]{charts/compare3Correctness.png}
    %     %\vspace{-0.1in}
    %     \caption{Comparing results of three variants.}
    %     %\vspace{-0.2in}
    %     \label{fig:compare3Correctness}
    % \end{subfigure}
    \includegraphics[width=0.45\textwidth]{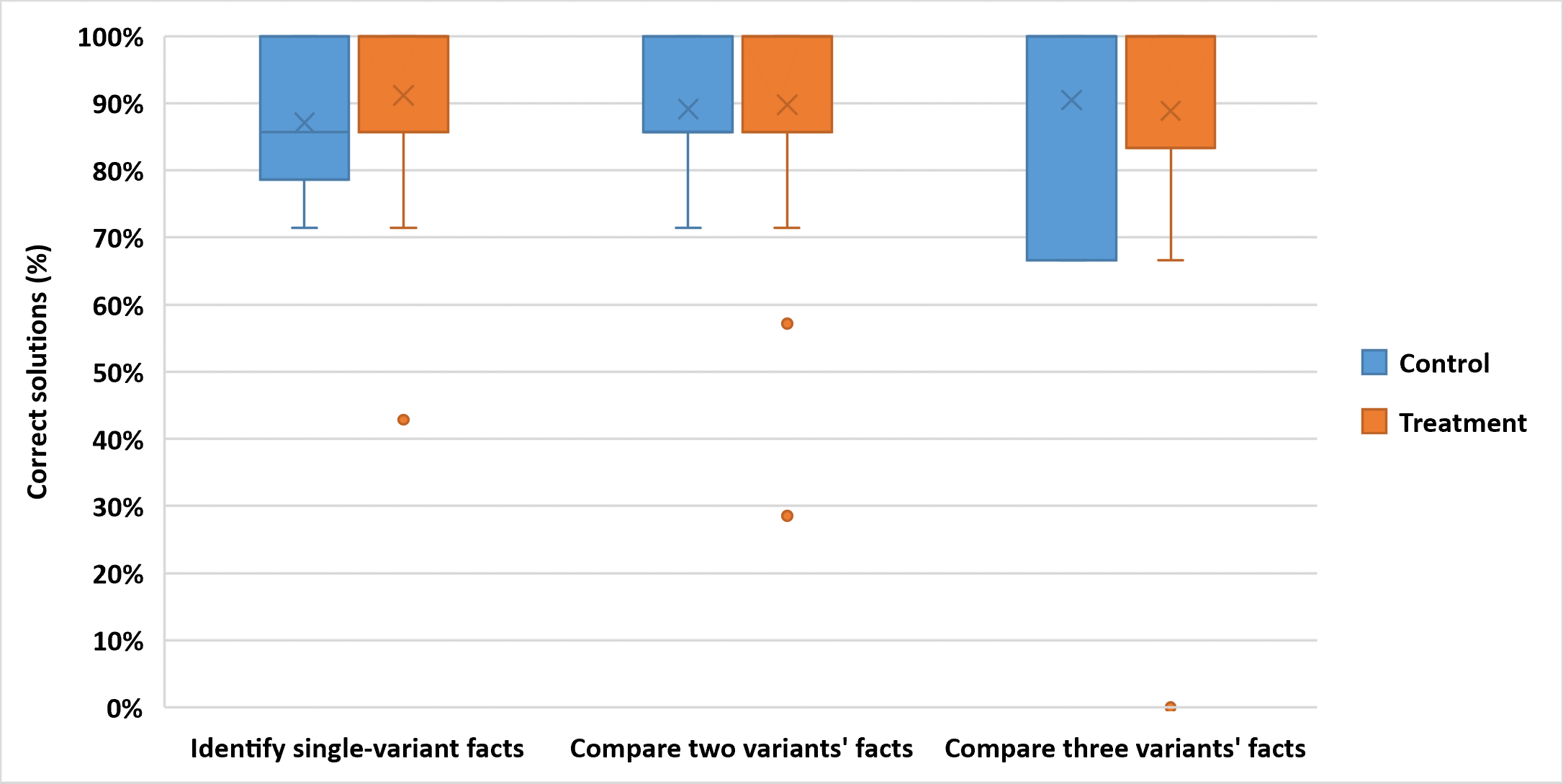}
%    \vspace{-0.1in}
    \caption{Average participant correctness, disaggregated by task category.}
    \label{fig:tasksCorrectness}
\end{figure}

Comparing the groups' accuracy disaggregating by task category (shown in Figure~\ref{fig:tasksCorrectness}), we observed no significant difference in the groups' accuracy in any task category.%TODO: plot and test are not converging (test exclude participants who did not use filters, plot does not exclude them)

% \begin{figure}[t]
% 	\centering
% 	\includegraphics[width=0.48\textwidth]{charts/identifyCorrectness.png}
% 	%\vspace{-0.1in}
% 	\caption{Correctness for tasks identifying a fact variant}
% 	%\vspace{-0.2in}
% 	\label{fig:identifyCorrectness}
% \end{figure}

% \begin{figure}[t]
% 	\centering
% 	\includegraphics[width=0.48\textwidth]{charts/compare2Correctness.png}
% 	%\vspace{-0.1in}
% 	\caption{Correctness for tasks comparing facts of two variants}
% 	%\vspace{-0.2in}
% 	\label{fig:compare2Correctness}
% \end{figure}

% \begin{figure}[t]
% 	\centering
% 	\includegraphics[width=0.48\textwidth]{charts/compare3Correctness.png}
% 	%\vspace{-0.1in}
% 	\caption{Correctness for tasks comparing facts of three variants}
% 	%\vspace{-0.2in}
% 	\label{fig:compare3Correctness}
% \end{figure}

\subsection{Cognitive Load}

Our null hypothesis related to cognitive load was:

\begin{enumerate}[label=H\arabic*]
\setcounter{enumi}{2}
\item \textbf{Cognitive load:} Using coloured filters does not change the cognitive load of performing comprehension tasks involving variability-aware analysis results.
\end{enumerate}

\begin{figure}[t]
    % \begin{subfigure}[]{0.33\textwidth}
    % 	% \centering
    % 	\includegraphics[width=\textwidth]{charts/cognitiveIdentify.png}
    % 	%\vspace{-0.1in}
    % 	\caption{Identifying a single variant's results.}
    % 	%\vspace{-0.2in}
    % 	\label{fig:cognitiveIdentify}
    % \end{subfigure}
    % % \hfill
    % \begin{subfigure}[]{0.33\textwidth}
    %     % \centering
    %     \includegraphics[width=\textwidth]{charts/cognitiveCompare2.png}
    % 	%\vspace{-0.1in}
    % 	\caption{Comparing results of two variants.}
    % 	%\vspace{-0.2in}
    % 	\label{fig:cognitiveCompare2}
    % \end{subfigure}
    % % \hfill
    % \begin{subfigure}[]{0.33\textwidth}
    %     % \centering
    %     \includegraphics[width=\textwidth]{charts/cognitiveCompare3.png}
    % 	%\vspace{-0.1in}
    % 	\caption{Comparing results of three variants.}
    % 	%\vspace{-0.2in}
    % 	\label{fig:cognitiveCompare3}
    % \end{subfigure}
    \includegraphics[width=0.45\textwidth]{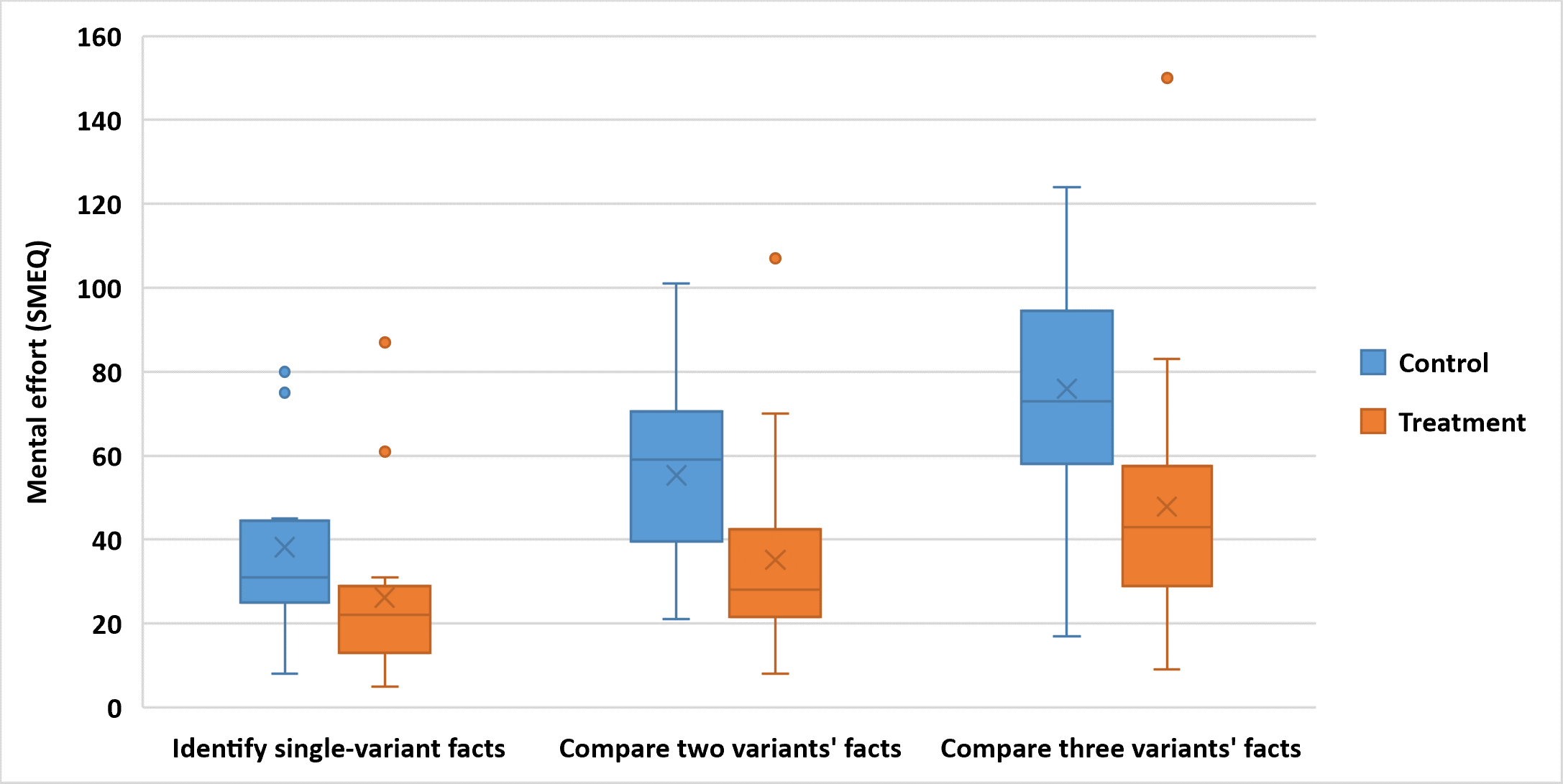}
%    \vspace{-0.1in}
    \caption{Average mental effort expended on tasks, disaggregated by task category.}
    \label{fig:tasksCognitive}
\end{figure}

\noindent This hypothesis is partially rejected, because  the results are significant for tasks comparing two 
%($p < 0.01$) 
or more variants ($p < 0.01$) but not as significant for tasks identifying results in a single variant ($0.05 < p < 0.1$). Figure~\ref{fig:tasksCognitive} shows the average cognitive loads reported by the participants, disaggregated by task category. Treatment participants indicated that identifying results belonging to a single variant required more or less \textit{``A Little Effort''} and control participants indicated such tasks required \textit{``Some Effort''}, based on the SMEQ scale. Whereas, the treatment participants reported significantly lower levels of mental effort (36\%) than the control participants when performing tasks involving two or more variants. The extra effort expended by control participants is due to their needing to calculate the satisfiability of variant conditions manually.

One treatment participant classified the mental effort for all task categories as considerably high. This participant raced through the tutorial on filters, did not use filters in the first stage of the study (although they were placed in the treatment group), and among all participants had the worst performance with respect to accuracy. In the user preferences exercise, they strongly preferred the uncoloured option for visualizing variants of interest.
%and their classification of the mental effort also expresses their preference not to use the coloured edges. %Section~\ref{sec:discussion} elaborates more on the rationale of that and other outlier participants. 

% \begin{figure}[t]
% 	\centering
% 	\includegraphics[width=0.48\textwidth]{charts/cognitiveIdentify.png}
% 	%\vspace{-0.1in}
% 	\caption{Mental effort for tasks identifying a fact variant}
% 	%\vspace{-0.2in}
% 	\label{fig:cognitiveIdentify}
% \end{figure}

% \begin{figure}[t]
% 	\centering
% 	\includegraphics[width=0.48\textwidth]{charts/cognitiveCompare2.png}
% 	%\vspace{-0.1in}
% 	\caption{Mental effort for tasks comparing facts of two variants}
% 	%\vspace{-0.2in}
% 	\label{fig:cognitiveCompare2}
% \end{figure}

% \begin{figure}[t]
% 	\centering
% 	\includegraphics[width=0.48\textwidth]{charts/cognitiveCompare3.png}
% 	%\vspace{-0.1in}
% 	\caption{Mental effort for tasks comparing facts of three variants}
% 	%\vspace{-0.2in}
% 	\label{fig:cognitiveCompare3}
% \end{figure}

\subsection{User Preferences}

Our null hypothesis related to user preference was:

\begin{enumerate}[label=H\arabic*]
\setcounter{enumi}{3}
\item \textbf{User preference:} Users feel indifferent about using coloured filters to highlight variability-aware analysis results.  
\end{enumerate}

\begin{figure}[t]
    \begin{subfigure}[]{0.45\textwidth}
    	% \centering
    	\includegraphics[width=\textwidth]{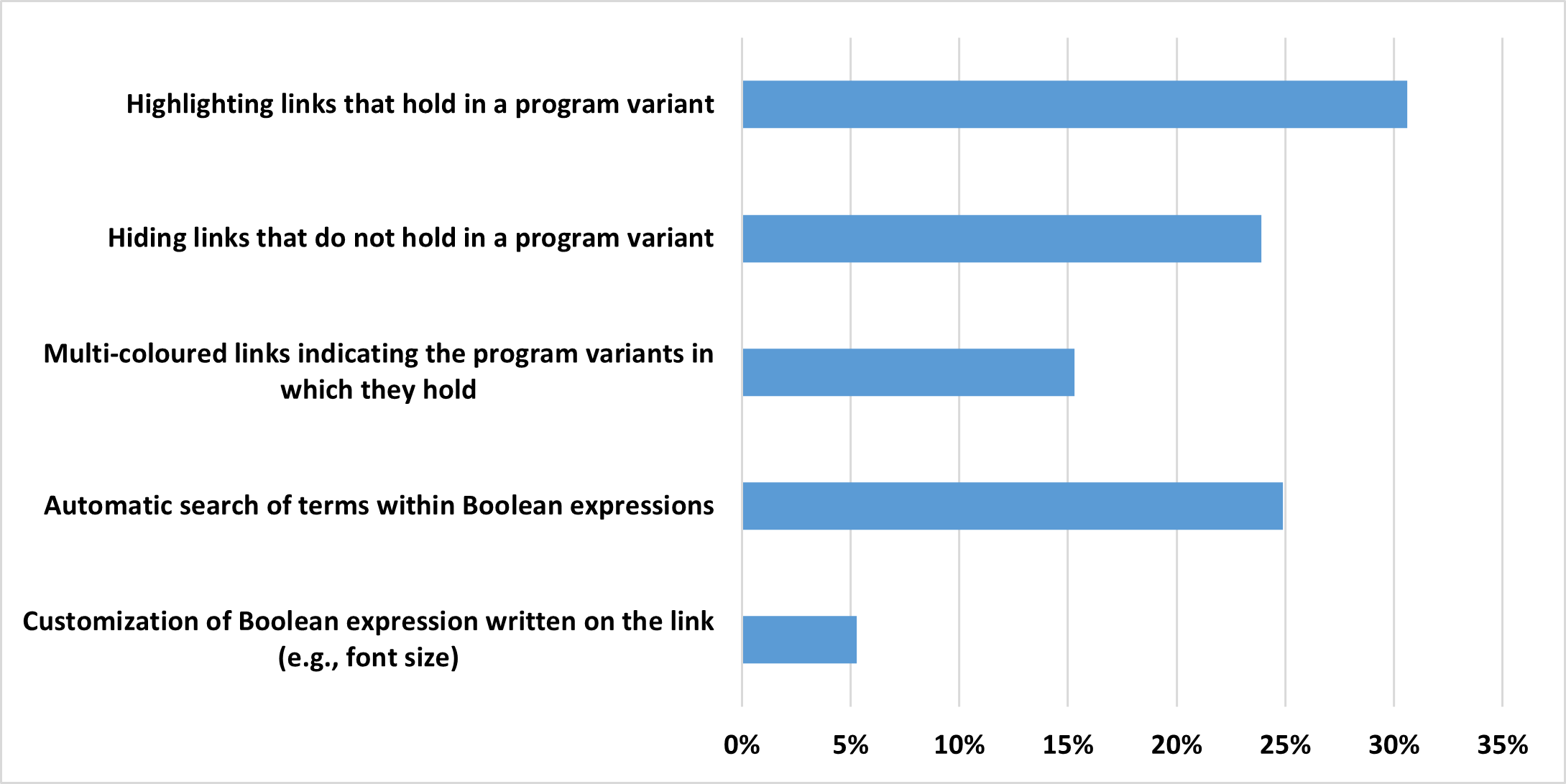}
    	% \vspace{-0.1in}
    	\caption{Control group's preferred choices for improvements to visualizer.}
    	%\vspace{-0.2in}
    	\label{fig:controlFeaturePreference}
    \end{subfigure}

    \begin{subfigure}[]{0.45\textwidth}
        % \centering
        \includegraphics[width=\textwidth]{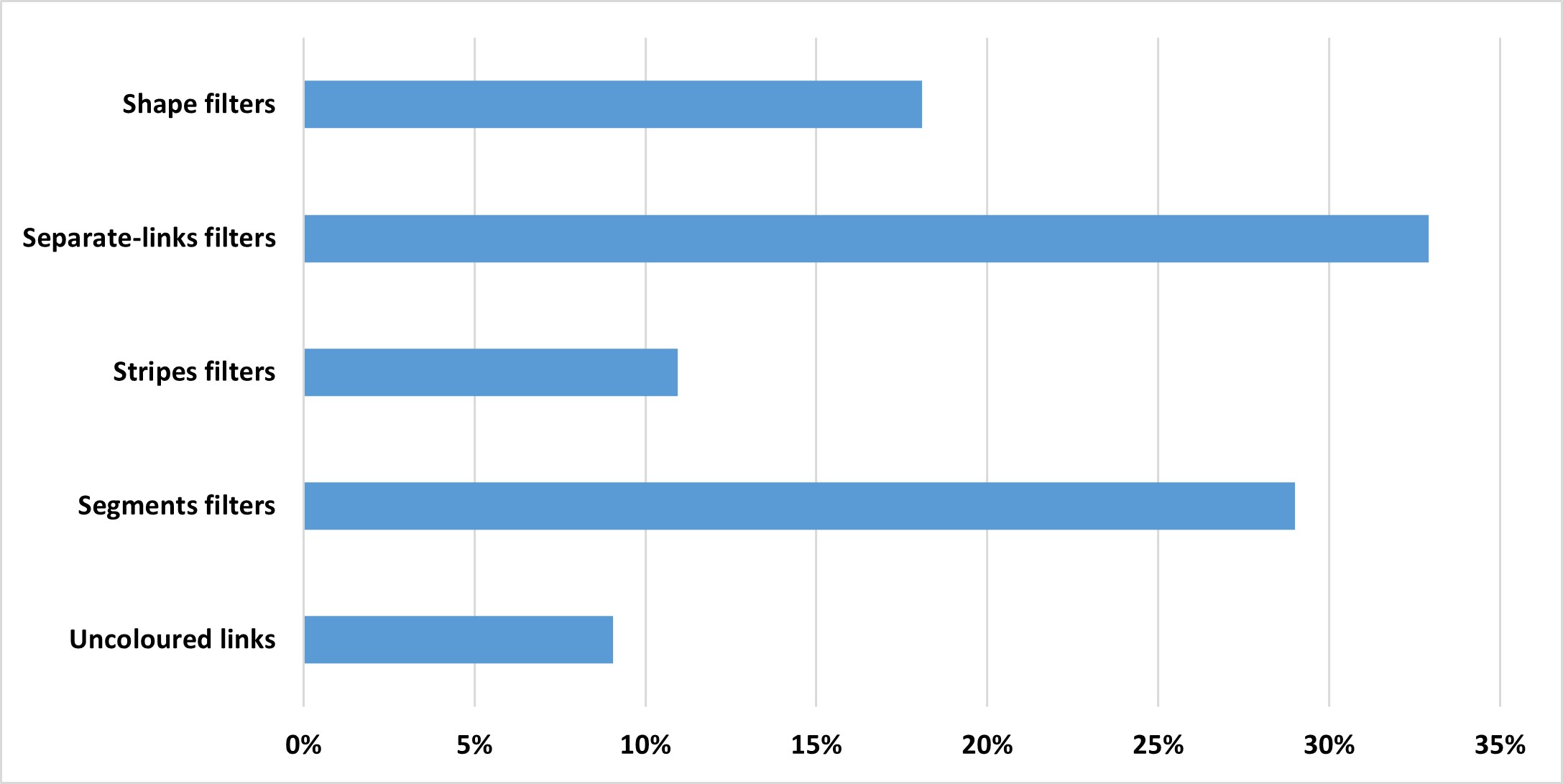}
    	%\vspace{-0.1in}
    	\caption{Treatment group's preferred choices for visualizing variant-specific results.}
    	%\vspace{-0.2in}
    	\label{fig:layoutPreference}
    \end{subfigure}
%    \vspace{-0.1in}
    \caption{Participants' preferred features}
    \label{fig:stagesPreference}
\end{figure}

\noindent This hypothesis is rejected ($p < 0.01$). Figure~\ref{fig:stagesPreference} shows the ranked preferences of visualization features, disaggregated by study group. Both participant groups expressed strong preferences for having coloured filters. Among control group participants, 31\% ranked \textit{``Highlighting links that hold in a program variant''} as their favourite feature for an improved interface. Among the treatment participants, only 9\% favoured uncoloured models. In fact, 71\% of the treatment participants distributed \textit{none} of their \$100 ``votes'' to the uncoloured filter options. 
% followed by "Automatic search of terms within Boolean expressions" (25\%), "Hiding links that do not hold in a program variant"(24\%), "Multi-coloured links indicating the program variants in which they hold"(15\%), and "Customization of Boolean expression written on the link (e.g., font size)"(5\%).

% \begin{figure}[t]
% 	\centering
% 	\includegraphics[width=0.48\textwidth]{charts/controlFeaturePreference.png}
% 	%\vspace{-0.1in}
% 	\caption{Favourite features for an improved interface as voted by the control group}
% 	%\vspace{-0.2in}
% 	\label{fig:controlFeaturePreference}
% \end{figure}

% \begin{figure}[t]
% 	\centering
% 	\includegraphics[width=0.48\textwidth]{charts/layoutPreference.png}
% 	%\vspace{-0.1in}
% 	\caption{Favourite edge-group visualization as voted by the control group}
% 	%\vspace{-0.2in}
% 	\label{fig:layoutPreference}
% \end{figure}

\subsection{Filter Visualization Options}

Our null hypothesis on differences between options for visualizing variant-specific results was:

\begin{enumerate}[label=H\arabic*]
\setcounter{enumi}{4}
\item\textbf{Filter Visualization Options:} The filters visualization options have the same impact on user efficiency and correctness.
\end{enumerate}

\begin{figure}[t]
    \begin{subfigure}[]{0.45\textwidth}
    	% \centering
    	\includegraphics[width=\textwidth]{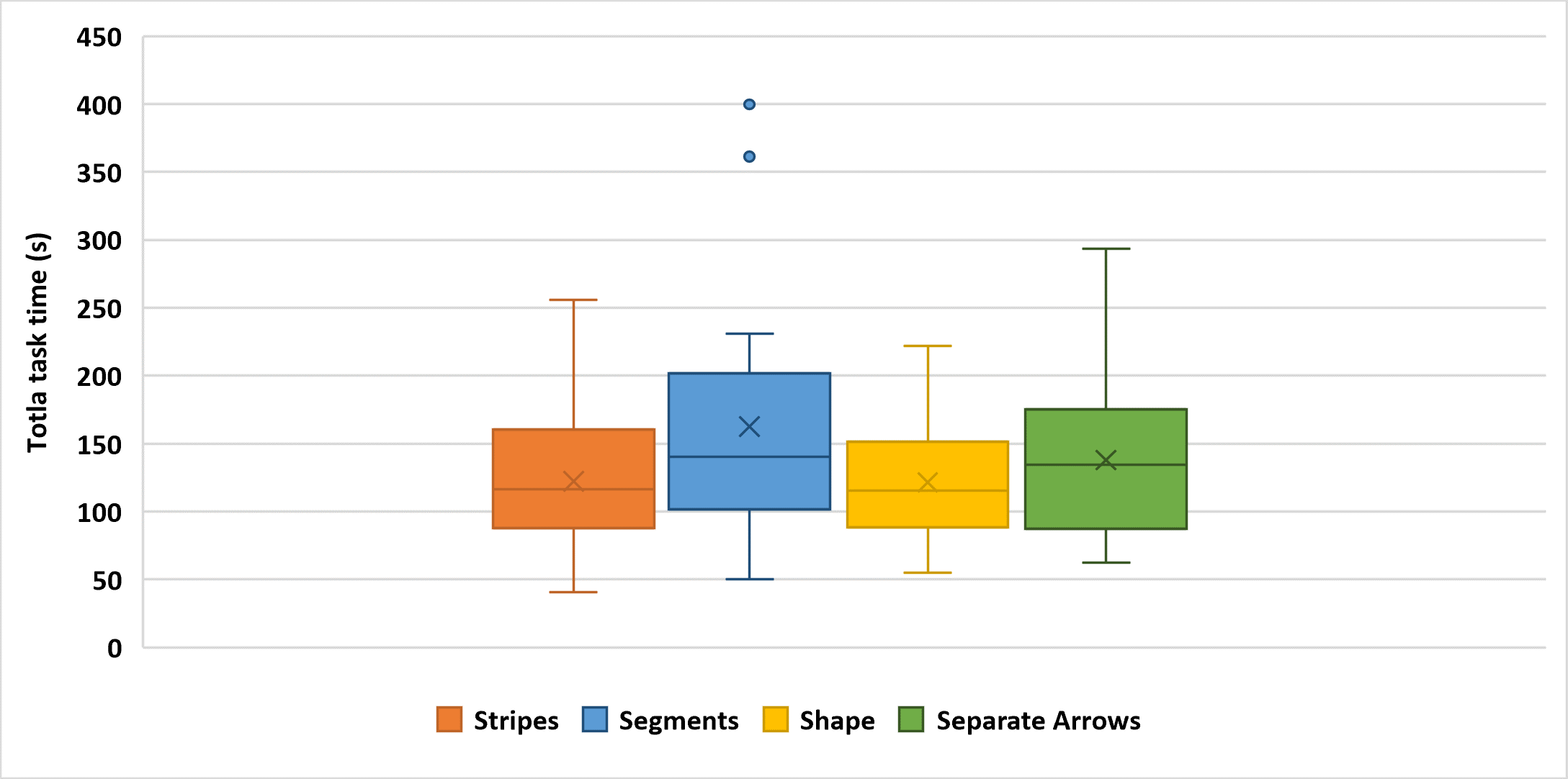}
    	%\vspace{-0.1in}
    	\caption{Time to complete second-stage tasks, disaggregated by filter visualization (treatment group only).}
    	%\vspace{-0.2in}
    	\label{fig:optionsTime}
    \end{subfigure}

    \begin{subfigure}[]{0.45\textwidth}
        % \centering
        \includegraphics[width=\textwidth]{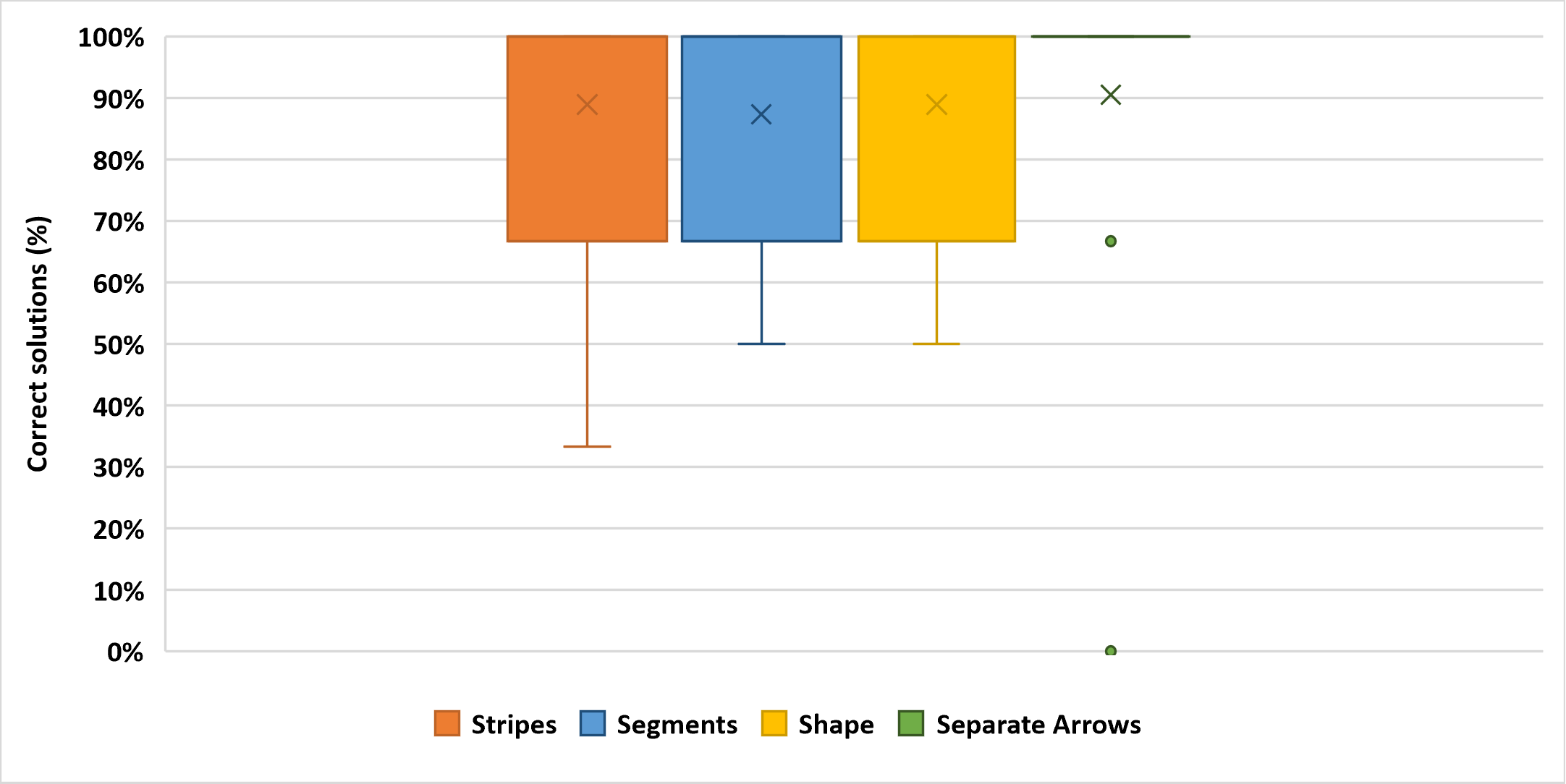}
    	%\vspace{-0.1in}
    	\caption{Correctness of second stage tasks, disaggregated by filter visualization (treatment group only).}
    	%\vspace{-0.2in}
    	\label{fig:optionsCorrectness}
    \end{subfigure}
 %   \vspace{-0.1in}
    \caption{Correctness and time comparison for filter visualization options.}
    \label{fig:optionsComparison}
\end{figure}

\noindent This hypothesis is accepted, as no difference was observed in the treatment participants' performance when using the different filter visualization options (considering a significance level of 1\%). Figure~\ref{fig:optionsComparison} compares the time and correctness of responses achieved by treatment participants using each of the filter visualization options. Even though more participants correctly solved all the tasks using the \textit{Separate Arrows} visualization, the average correctness for that visualization was not significantly different from the results achieved with the other options. As Figure~\ref{fig:optionsCorrectness} shows, few participants made mistakes using the \textit{Separate Arrows} visualization (4 out of 21). However, one specific outlier solved all tasks incorrectly with that visualization, which decreased significantly the average correctness for the \textit{Separate Arrows} visualization. 

%%%%%%%%%%%%%%
% TODO: remove control 

% \begin{figure}[t]
% 	\centering
% 	\includegraphics[width=0.48\textwidth]{charts/timeLayout.png}
% 	%\vspace{-0.1in}
% 	\caption{Time }
% 	%\vspace{-0.2in}
% 	\label{fig:timeLayout}
% \end{figure}

% \begin{figure}[t]
% 	\centering
% 	\includegraphics[width=0.48\textwidth]{charts/correctnessLayouts.png}
% 	%\vspace{-0.1in}
% 	\caption{}
% 	%\vspace{-0.2in}
% 	\label{fig:correctnessLayouts}
% \end{figure}

Table~\ref{tab:ptests} summarizes the results of our study's null-hypothesis tests, including results disaggregated by the stage of the study and the task category. Rejected null hypotheses are highlighted in bold.

\subsection{Participants' Feedback}

\begin{figure}[t]
    \begin{subfigure}[]{0.45\textwidth}
    	% \centering
    	\includegraphics[width=\textwidth]{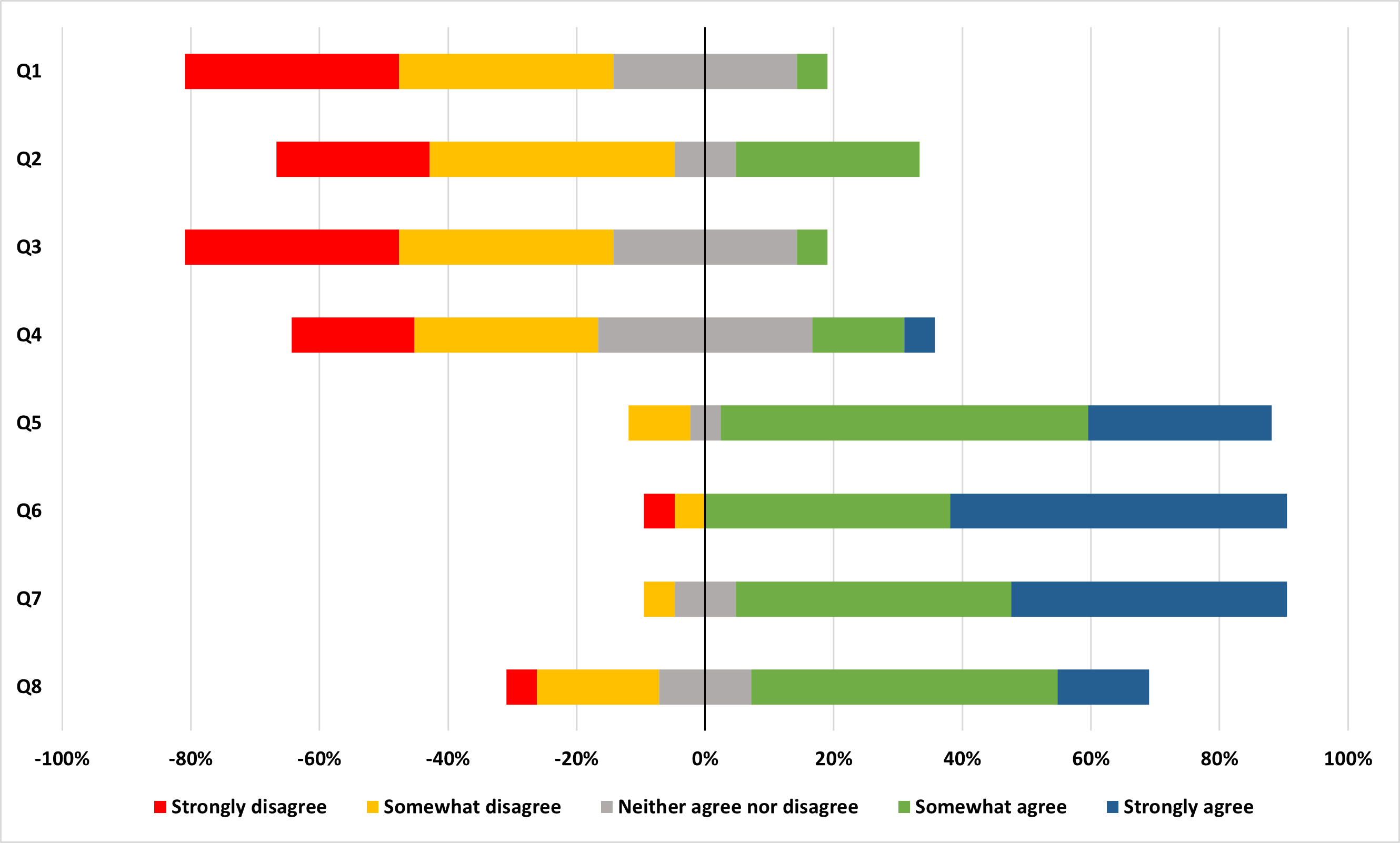}
    	%\vspace{-0.1in}
    	\caption{Control group}
    	%\vspace{-0.2in}
    	\label{fig:controlStatements}
    \end{subfigure}

    \begin{subfigure}[]{0.45\textwidth}
        % \centering
        \includegraphics[width=\textwidth]{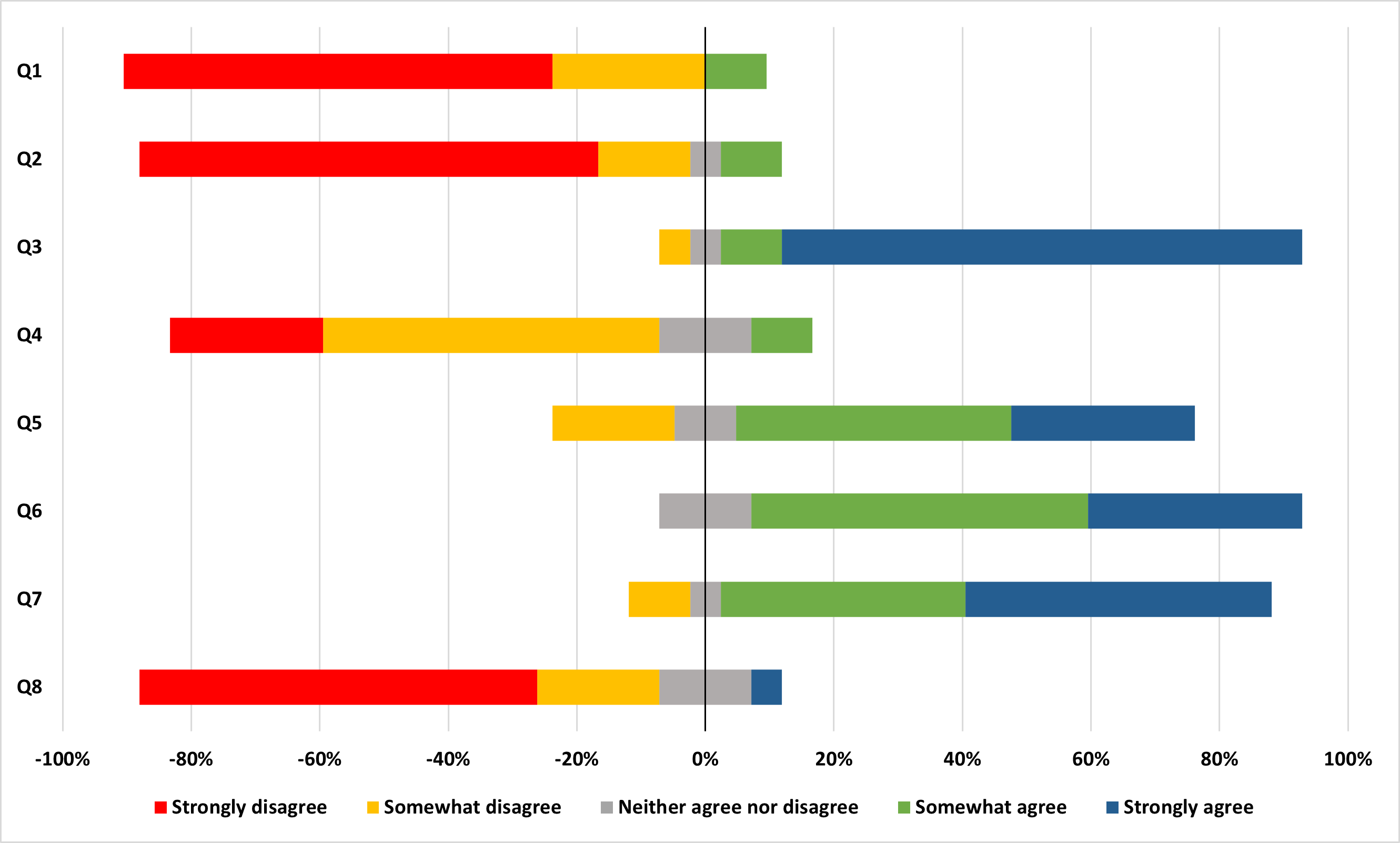}
    	%\vspace{-0.1in}
    	\caption{Treatment group}
    	%\vspace{-0.2in}
    	\label{fig:treatmentStatements}
    \end{subfigure}
%    \vspace{-0.1in}
    \caption{Level of agreement to experience statements.}
    \label{fig:statementsResults}
\end{figure}

Participants' responses to the list of statements and the open-ended questions helped us understand their opinions about the coloured filters. Figure~\ref{fig:statementsResults} shows the agreement levels of study participants to the presented statements. The treatment group is more homogeneous in strongly disagreeing that coloured filters are complex and cumbersome (Q2). When shown the same statement about Boolean expressions, the control group gave more diverse responses, with more participants being neutral or slightly agreeing with the statement. 

Control participants also acknowledge that they need to expend extra effort when dealing with Boolean expressions and uncoloured models. More than 50\% of the control participants either agreed or were neutral about saying there was a significant learning curve before they could identify the program variants in which particular results hold (Q4). Fewer than a quarter of the treatment participants agreed with that statement. That said, 85\% of the control participants agreed that Boolean expressions are easy to access (Q5), whereas 71\% of the treatment group says the same about the filter customization.

% \begin{figure}[t]
% 	\centering
% 	\includegraphics[width=0.48\textwidth]{charts/controlStatements.png}
% 	%\vspace{-0.1in}
% 	\caption{Control group statements}
% 	%\vspace{-0.2in}
% 	\label{fig:controlStatements}
% \end{figure}

% \begin{figure}[t]
% 	\centering
% 	\includegraphics[width=0.48\textwidth]{charts/treatmentStatements.png}
% 	%\vspace{-0.1in}
% 	\caption{Treatment group statements}
% 	%\vspace{-0.2in}
% 	\label{fig:treatmentStatements}
% \end{figure}

Participants' feedback collected by the open-ended questions highlighted the strengths and weaknesses of the visualizer. Overall, all participants considered the visualizer easy to use and intuitive. Some of the participants acknowledged minor problems they had with the visualizer, such as arrowhead sizes, text styling, and unclear colour selection. Treatment participants who chose not to use the filters in the first stage of the study singled out the difficulty in selecting the colour as the main obstacle to using the filters. Other improvements suggested by the participants include making the filter customization easier to use and improving the presentation of Boolean expressions. 

Lastly, participants in both groups raised concerns with the visualization of large models. In particular, they worried that their presentation would be cluttered and that textual labels would overlap making them difficult to read. Although the model layout was not assessed as part of this study, this concern is worth investigating in the future.

\begin{table}[]
\caption{Statistical Tests of Hypotheses (rejected hypotheses in bold).}\label{tab:ptests}

\resizebox{0.49\textwidth}{!}{%
\begin{tabular}{|ll|c|}
\hline
\multicolumn{1}{|c|}{\textbf{Hypothesis}}                                                                         & \textbf{Sub-Hypothesis}                & \textbf{P-Value}    \\ \hline
\multicolumn{1}{|c|}{}                                                                                            & \textbf{Stage 1}                       & $\boldsymbol{p < 0.01}$ \\ \cline{2-3} 
\multicolumn{1}{|l|}{}                                                                                            & \textbf{Stage 2}                       & $\boldsymbol{p < 0.01}$ \\ \cline{2-3} 
\multicolumn{1}{|c|}{\textbf{H1. Efficiency}}                                                                     & \textbf{Identify variants' facts}      & $\boldsymbol{p < 0.01}$ \\ \cline{2-3} 
\multicolumn{1}{|l|}{}                                                                                            & Compare two variants' facts            & $0.05 < p < 0.1$    \\ \cline{2-3} 
\multicolumn{1}{|l|}{}                                                                                            & \textbf{Compare three variants' facts} & $\boldsymbol{p < 0.01}$ \\ \hline
\multicolumn{1}{|l|}{}                                                                                            & \textbf{Stage 1}                       & $\boldsymbol{p < 0.01}$ \\ \cline{2-3} 
\multicolumn{1}{|l|}{}                                                                                            & Stage 2                                & $0.2 < p$           \\ \cline{2-3} 
\multicolumn{1}{|c|}{\textbf{H2. Correctness}}                                                                    & Identify variants' facts               & $0.2 < p$           \\ \cline{2-3} 
\multicolumn{1}{|l|}{}                                                                                            & Compare two variants' facts            & $0.2 < p$           \\ \cline{2-3} 
\multicolumn{1}{|l|}{}                                                                                            & Compare three variants' facts          & $0.2 < p$           \\ \hline
\multicolumn{1}{|l|}{}                                                                                            & Identify variants' facts               & $0.05 < p < 0.1$    \\ \cline{2-3} 
\multicolumn{1}{|c|}{\textbf{H3. Cognitive Load}}                                                                 & \textbf{Compare two variants' facts}   & $\boldsymbol{p < 0.01}$ \\ \cline{2-3} 
\multicolumn{1}{|l|}{}                                                                                            & \textbf{Compare three variants' facts} & $\boldsymbol{p < 0.01}$ \\ \hline
{\textbf{H4. User Preference}} &                                                                                                        & $\boldsymbol{p < 0.01}$ \\ \hline
%\multicolumn{2}{|c|}{\textbf{H4. User Preference}}                                                                                                         & $\boldsymbol{p < 0.01}$ \\ \hline
\multicolumn{1}{|c|}{\multirow{2}{*}{\begin{tabular}[c]{@{}c@{}}H5. Filter Visualization\\ Options\end{tabular}}} & Efficiency                             & $0.2 < p$           \\ \cline{2-3} 
\multicolumn{1}{|c|}{}                                                                                            & Correctness                            & $0.2 < p$           \\ \hline
\end{tabular}%
}
\end{table}

\section{Threats to validity}

Our user study may present threats to construct validity due to limitations of the study design and biases introduced by the student participants. We aimed to partially mitigate those threats by balancing out the study experience and ascertaining student proficiency with configurable software.

The level of proficiency of participants with respect to the development of configurable software development and the study design may have threats to external validity. We mitigated these threats by assessing the eligibility of participants before study participation and using a rich yet accessible standard product line~\cite{lopez2001standard} for the basis of the study. Moreover, our screening questionnaire helps eliminate the threat that participants have biases against reasoning about Boolean expressions.

% Internal validity
% Learning effects -> task order is constant; the treatment options are presented in a variety of orders 
% Unbalance participant allocation -> alternate assignment

% External validity
% Lack of familiarity of the participants -> screening questionnaire, prior working experience
% Presence conditions were given ->we believe, no effect
% No realistic or challenging tasks -> use of a standard problem, comprehension tasks
% Large systems -> future work and studying comprehension in large system is desirable but can be limiting in an experimental design

It is important to note that participants' performances in the study did not include the time it would take to formulate variants' presence conditions. In the study, both the control and the treatment groups were given the PC expressions that specify the variant(s) of interest for each task. 
% This means that the treatment group could simply specify filters by copying the provided presence condition and pasting it into the filter editor's textbox, and 90\% of the treatment participants did exactly that. 
We do not believe that providing these PC expressions affects the results of the study. In practice, an engineer would need a sufficiently precise specification of the variants of interest regardless of whether they are working with the original Neo4j Browser and are trying to filter analysis results manually or are working with the treatment visualizer and need to formulate filter expressions. The study evaluates only whether the treatment visualizer eases tasks related to reading, comprehending, and reasoning about variability-aware analysis results - it does not evaluate the user's experience in formulating variants' PCs, which remains a challenging and error-prone task.

An obvious threat to validity is the generalizability of our results to large SPLs that have more program data and variants. Mitigating this threat is not a simple challenge, as program data limits on the amount of information that user study participants are capable of consuming before becoming fatigued. We partially mitigate the threat by using a manageable but reasonably diverse SPL as the basis of our user study.

%The learning effects of the interface and unbalanced allocation of participants could also pose threats to internal validity. To mitigate the effect of such threats, we present a constant order of tasks to all participants. For the treatment group, we also varied the order in which the treatment visualization options were used to generate enough data points to support statistically valid findings. The allocation of participants was alternated to ensure that both control and treatment groups had the same number of participants.

The learning effects of the interface and the unbalanced allocation of participants could also pose threats to internal validity. To mitigate the effect of such threats, we present a constant order of tasks to all participants. For the treatment group, we also varied the order in which the treatment visualization options were used to generate enough data points to support statistically valid findings. The allocation of participants was alternated to ensure that both control and treatment groups had the same number of participants.

\section{Discussion}
\label{sec:discussion}

% Expected results

% \begin{itemize}
%     \item Filters are faster and mentally cheaper => \forme{Looking at colour is (timely and cognitively) better than mentally calculating Boolean expression. Removing the expressions and relying on the filters may help clean up the graph.}

%      \item Filters increase correctness when the user is able to create and customize their filters => \forme{The cognitive gains of creating and customizing filters help users to make fewer mistakes. We should improve the interface to meet the user's needs and facilitate personalizing the visualization.}
    
%     \item Users prefer some of the filter visualization options over others => \forme{Some of the visualization options are considered irrelevant or unuseful. We should revise the visualization options offered (or how we offer them).}

% \end{itemize}

% Unexpected results

% \begin{itemize}
%     \item Filters do not increase correctness when the user is given models with pre-formatted filters. => \forme{The personal preferences and predictability seem to influence the performance of the user. Making a transparent and versatile interface should help users achieve better results.}
    
%     \item Filter visualization options do not have a significant difference in performance => \forme{Since filter visualization options deliver similar results, it would be good to facilitate the use of popular visualization options and remove less popular visualization options}
% \end{itemize}

% Discuss results that we expected
% Filters are faster
% Filters are cognitively cheaper
The results of our study suggest that 
%We learned that 
highlighting variant-specific model elements enables users to comprehend variability-aware analysis results in less time and with less mental effort. This result reflects also participants' descriptions of their experiences. Control participants recognized the difficulties the tasks (e.g., \textit{"...I still need to take quite some effort to verify the Boolean expression in my head, which is exhaustive."}), while some of the treatment participants highlighted the benefits of having access to the filters (e.g., \textit{"It greatly reduces the mental load by allowing you to abstract the program variant without considering every parameter."}). The treatment participants completed their tasks more quickly despite the extra work of creating and customizing filters. The reduced mental effort was especially evident when tasks compare the analysis results for two or more program variants. We believe that such gains could further improved if
%motivate future versions of 
the visualizer allowed users to choose whether the PC labels on model elements are elided or exposed.

% Discuss results that we did not expect
% Mixed results for correctness comparison (first stage is better, second stage is not)
% Explain types of errors
The impact of coloured filters on task correctness depended on participants' interactions with the models. When given the option to customize the models, the participants who used coloured filters performed significantly better than those who were forced to interpret the model's Boolean expressions. Two of the most common errors made by the control group were (1) misreading the direction of edges (i.e., arrowheads) and (2) reporting \textit{paths} that originate from a particular node instead of reporting only the node's direct neighbours. We hypothesize that the high mental effort and attention needed to reason about the model's presence conditions and customize the model (e.g., move nodes around, hide edges of a particular type) distracted them from fundamental details of the model and the task. 

% We modified the interface to reduce the chances of those errors. First, we highlighted parts of the text that specified the requirement of directly connected nodes. Also, after realizing the difficulties the first ten participants had in reading the arrows' direction, we increased the default size of the arrowhead and added a customization option to allow users to set their preferred size. However, participants of the control group still made such mistakes quite frequently. 

When participants did not have the freedom to customize the models, the coloured filters did not have as significant an impact on the correctness of the participants' work. 
%This difference in impact is due to both the improved performance of the control group and the more diverse performance of the treatment group. 
The control participants seem to benefit from being unable to customize the models. Rather than being subjected to the tool's automated layout of models, they were provided readable and uncluttered models, which lowered the required mental effort and allowed them to produce more accurate results compared with their performance in the first stage of the study. Also, the models provided in the second stage had a consistent look in every task, which might have increased familiarity with the models, leading to improve performance. 

The treatment participants were also given readable and uncluttered models in the second stage of the study, but different tasks employed different filter visualizations, and the changing visualizations may have affected their second-stage performance. The different filter visualization options resulted in no significant difference in participants' time or correctness of task results, but participants had personal preferences.
%among the visualization options. 
Two options, \textit{Separate Arrows} and \textit{Arrow Segments}, received more than half the participants' ``votes''. 
%We believe that forcing participants to use less favoured visualizations of variants and removing the capacity to customize the visualization affected the performance of some participants, increasing the variability of the results with respect to correctness. 
The participants' feedback suggest that the interface could be improved by merging the \textit{Arrow Shapes} and \textit{Arrow Segments} options (since a solid segment is a type of shape) and removing the least desired \textit{Arrow Stripes} option.

%%%%%%%%%%%%%%%%%%%%%%%%%%
% What are the differences in time and correctness for different types of tasks?
%%%%%%%%%%%%%%%%%%%%%%%%%%

The two stages of the study had the unintended effect of assessing in the first stage participants' \textit{realistic} experiences in using the visualizer (i.e., its automated layouts of query results and its filter creation interface) and assessing in the second stage a more \textit{idealized} experience, where pre-formatted models were nicely laid out.
%and, for the treatment group, filter visualization options were pre-defined.
Thus, participants' feedback on the provided ``idealized'' models reveal more fundamental issues with the two visualizers, such as cluttered graphs and overlapping labels. For example, 
%when asked about problems they faced with the visualizer, 
one participant  mentioned \textit{"Cluttered graphs need manual organization"} and another said \textit{"Minor UI issues that bothered me such as text covering some other text and having to zoom out a lot to see the whole graph"}. The control participants' rankings of desired visualizer features favoured being able to search for terms used in Boolean expressions and to hide edges that do not hold in a program variant. 
%Supporting these user requests would involve developing manageable querying and incremental exploration of relevant parts of the model without introducing confusion due to the eliding of information. 
The development of such design choices should be the future focus of the visualizer.

% Discuss preference and cognitive load from P30 

% Add stuff that we think based on the results - control group are consumed by the task of customizing
% Error types for correctness - it tells where their attention was
% Delve more deeply into the comments people make

% If it is too much -> select and hide, but still have to provide big picture. We provided the ability to hide but not everyone used it
% Lessons learned - make 

% Compare correctness and time for different types of tasks
% Write like we are not author of the UI, the UI was done in ESME and we got the software it was poorly evaluated (they only commented on the provided images)

\section{Background and Related Works}
\label{sec:background}

Our work differs from previous works in its focus on the evaluation of visualizing variability-aware analysis results and the effectiveness of the visualization options.
%evaluating the utility of software visualization in both the data needs and the task goals determining the suitability of the tool. Works evaluating analysis-results visualization tools focus on simpler data encodings, program analyses, and user tasks during their evaluation.

\subsection{Visualizing Program-Analysis Results}

The conventional use of graphical software models is to support the analysis and understanding of relations between software elements. Such understanding
%relations provide an overview of the software structure that 
can facilitate several software processes, such as software evolution~\cite{wilde2018merge}, software refactoring~\cite{beck2011visual}, and reverse engineering~\cite{storey2001shrimp, daniel2014polyptychon}. More specifically, lower-level maintenance tasks~\cite{sillito2008asking} and the reachability questions~\cite{latoza2010developers} posed by engineers while performing such tasks are among the primary goals of interactive graphical software tools. 

For example, IDE tools like Zest Sequence Viewer~\cite{bennett2007working}, Graph Buddy~\cite{borowski2022graph}, Reacher~\cite{latoza2011visualizing}, and ReachHover~\cite{yoo2022investigating} enable the exploration and visualization of different types of connective program data (e.g., control flow traces, dataflow traces, dependencies, inheritance). The user selects the types of code relations and elements of interest to initiate an interactive exploration of the program model. Similarly, JavaRelationshipGraphs~\cite{arora2019javarelationshipgraphs} extracts a graphical model of a program and uses the Neo4j database and browser for further analyses of the software, using colour to represent different types of relations.
% Unanonymized version
% The interface evaluated in this work differs from previous visualizers as its software graphical models focus on supporting the comprehension of variability-aware analysis results. When compared to previous works' models, the evaluated graphs include an extra layer of information which adds the presence conditions of the code relations. Such data extends the scope of supported questions by adding a variability aspect to the reachability questions supported by the cited tools. Also, adding the coloured filters to Neo4j Browser leverages graph visualization techniques used to represent graph group structures~\cite{vehlow2017visualizing}.
The comprehension tasks and research questions in this study are different from those used in previous tool evaluations as they focus on understanding and reasoning about variability-aware analysis results. 
%The data consumption evaluated in this study extends the scope of tool-supported questions by adding a variability aspect to the reachability questions supported by the cited tools. 
%Also, our work aims to evaluate a tool more completely, as some of the previous tools lack more mature user studies with a higher number of participants and realistic diversity of use scenarios. 

\subsection{Visualizing Software Product Lines}

Previous works that use colour in SPL visualization do so to distinguish among features or variability points in the SPL's source code or feature model. Tools like CIDE~\cite{kastner2009guaranteeing}, FeatureMapper~\cite{Heidenreich-VISPLE08}, fmp2rsm~\cite{czarnecki2006verifying}, and FeatureVISU~\cite{Apel:2011} colourize model or source-code elements according to their association with the features that the user selects. Such visualization tools contribute to the engineer's comprehension and productivity when modifying feature configurations~\cite{asadi2016effects}.

Other works use coloured models to visualize the results of specific types of SPL analyses, such as identifying obsolete features~\cite{loesch2007optimization}, consequences of configuration decisions~\cite{botterweck2008visual}, and feature constraints~\cite{martinez2014feature}. In these cases, the colours highlight the model elements that pertain to each reported analysis result rather than highlighting the results that pertain to distinct product variants. The spatial distribution of the model elements is also explored by some of those visualizers. However, the cited works do not include observed experiments with users in their evaluation. They normally evaluate the correctness of their visualization through a case study. When user feedback is mentioned, it is unclear whether it is based on users' actual experience with the tool.
%the users performed any tasks with the method or not.

The evaluations performed by Str{\"u}ber et al.~\cite{struber2020variability} and Shahin et al.~\cite{shahin2023applying} are closest to our work, but the former diverges in the scope of the tool, and the latter lacks a proper user study. In~\cite{struber2020variability}, multiple variability representations are evaluated in a user study with many participants performing variability comprehension tasks that are similar to those used in our evaluation. Their work focuses on the visualization of variable class diagrams, while our work evaluates the comprehension of variability-aware analysis results. Shahin et al.~\cite{shahin2023applying} introduced the coloured filter interface evaluated in this work. However, their evaluation was limited to users' impressions and suggestions based on a demonstration of the tool. Our work evaluates the interface more thoroughly by measuring and analyzing the impact of the coloured filters on user understanding.
% Struber: evaluates comprehension of class diagrams

% The visualizer evaluated in our work uses colour to highlight model relations that belong to SPL variants specified by the engineer. The engineer provides the feature configurations representing different variants, and the visualizer paints the analysis results with the corresponding colours of all variants in which the relation exists. 

\section{Conclusion}

Comprehending SPL analysis results can be particularly challenging for engineers. On top of the difficulties of parsing and understanding reported program data, the user is expected to mentally determine whether results hold in  variants of interest. We have evaluated the use of an interactive visualizer that supports comprehension of variability-aware analysis results by highlighting those that belong to relevant variants specified by the user. We learned that the visualizer can provide significant gains in time, correctness, and cognitive load to users performing a variety of comprehension tasks. The study's results also suggest future improvements to the visualizer including layout algorithms that avoid overlapping model labels, optional eliding of PC annotations, and simplifications to the filter visualization options.

%In the future, we hope to extend the capacities of the visualizer and its evaluation to assess the impact the tool can have when working with large SPLs.

% \bibliographystyle{ACM-Reference-Format}
\bibliographystyle{IEEEtran}
\bibliography{bibliography}

% Generated by IEEEtran.bst, version: 1.14 (2015/08/26)
\begin{thebibliography}{10}
\providecommand{\url}[1]{#1}
\csname url@samestyle\endcsname
\providecommand{\newblock}{\relax}
\providecommand{\bibinfo}[2]{#2}
\providecommand{\BIBentrySTDinterwordspacing}{\spaceskip=0pt\relax}
\providecommand{\BIBentryALTinterwordstretchfactor}{4}
\providecommand{\BIBentryALTinterwordspacing}{\spaceskip=\fontdimen2\font plus
\BIBentryALTinterwordstretchfactor\fontdimen3\font minus
  \fontdimen4\font\relax}
\providecommand{\BIBforeignlanguage}[2]{{%
\expandafter\ifx\csname l@#1\endcsname\relax
\typeout{** WARNING: IEEEtran.bst: No hyphenation pattern has been}%
\typeout{** loaded for the language `#1'. Using the pattern for}%
\typeout{** the default language instead.}%
\else
\language=\csname l@#1\endcsname
\fi
#2}}
\providecommand{\BIBdecl}{\relax}
\BIBdecl

\bibitem{latoza2010developers}
T.~D. LaToza and B.~A. Myers, ``Developers ask reachability questions,'' in
  \emph{Proceedings of the ACM/IEEE International Conference on Software
  Engineering (ICSE'10)}, 2010, pp. 185--194.

\bibitem{sillito2008asking}
J.~Sillito, G.~C. Murphy, and K.~De~Volder, ``{Asking and Answering Questions
  during a Programming Change Task},'' \emph{IEEE Transactions on Software
  Engineering}, vol.~34, no.~4, pp. 434--451, 2008.

\bibitem{barnett2014get}
M.~Barnett, R.~DeLine, A.~Lal, and S.~Qadeer, ``Get me here: Using verification
  tools to answer developer questions,'' Technical Report MSR-TR-2014-10, Tech.
  Rep., 2014.

\bibitem{latoza2011}
{LaToza, Thomas D.}, ``{Answering Reachability Questions},'' Ph.D.
  dissertation, {Institute for Software Research, Carnegie Mellon University},
  2011.

\bibitem{yoo2022investigating}
J.~Yoo, ``{Investigating Data-flow Reachability Questions},'' Master's thesis,
  University of British Columbia, 2022.

\bibitem{Clements:2001}
P.~Clements and L.~Northrop, \emph{{Software Product Lines: Practices and
  Patterns}}.\hskip 1em plus 0.5em minus 0.4em\relax Addison-Wesley
  Professional, 2001.

\bibitem{spllift-PC}
E.~Bodden, T.~Tol{\^e}do, M.~Ribeiro, C.~Brabrand, P.~Borba, and M.~Mezini,
  ``{SPLLIFT: Statically Analyzing Software Product Lines in Minutes Instead of
  Years},'' \emph{ACM SIGPLAN Notices (SIGPLAN'13)}, vol.~48, no.~6, pp.
  355--364, 06 2013.

\bibitem{ModelChecking-PC}
A.~Classen, P.~Heymans, P.-Y. Schobbens, A.~Legay, and J.-F. Raskin, ``{Model
  Checking Lots of Systems: Efficient Verification of Temporal Properties in
  Software Product Lines},'' in \emph{Proceedings of the ACM/IEEE International
  Conference on Software Engineering (ICSE'10)}, 2010, pp. 335--344.

\bibitem{TypeSafety-PC}
S.~Apel, C.~K{\"a}stner, A.~Gr{\"o}{\ss}linger, and C.~Lengauer, ``{Type Safety
  for Feature-Oriented Product Lines},'' \emph{Automated Software Engineering},
  vol.~17, no.~3, pp. 251--300, 2010.

\bibitem{Analysis-PC}
J.~Liebig, A.~von Rhein, C.~K\"{a}stner, S.~Apel, J.~D\"{o}rre, and
  C.~Lengauer, ``{Scalable Analysis of Variable Software},'' in
  \emph{Proceedings of the Joint European Software Engineering Conference and
  Symposium on the Foundations of Software Engineering (ESEC/FSE'13)}, 2013, p.
  81–91.

\bibitem{ModelChecking-PC2}
K.~Lauenroth, K.~Pohl, and S.~Toehning, ``{Model Checking of Domain Artifacts
  in Product Line Engineering},'' in \emph{Proceedings of the IEEE/ACM
  International Conference on Automated Software Engineering (ASE'09)}, 2009,
  pp. 269--280.

\bibitem{shahin2023applying}
R.~Shahin, R.~Toledo, R.~Hackman, J.~M. Atlee, and M.~Chechik, ``{Applying
  Declarative Analysis to Industrial Automotive Software Product Line
  Models},'' \emph{Empirical Software Engineering}, vol.~28, no.~2, p.~40,
  2023.

\bibitem{Neo4j}
\BIBentryALTinterwordspacing
Neo4j, ``{Neo4j Graph Database},'' accessed July, 2023. [Online]. Available:
  \url{https://neo4j.com/}
\BIBentrySTDinterwordspacing

\bibitem{lopez2001standard}
R.~E. Lopez-Herrejon and D.~Batory, ``{A Standard Problem for Evaluating
  Product-line Methodologies},'' in \emph{Proceedings of the International
  Symposium on Generative and Component-Based Software Engineering (GCSE'01)},
  2001, pp. 10--24.

\bibitem{Kang:1990}
K.~Kang, S.~Cohen, J.~Hess, W.~Novak, and A.~Peterson, ``{Feature-Oriented
  Domain Analysis (FODA) Feasibility Study},'' Software Engineering Institute,
  Carnegie Mellon University, Tech. Rep. CMU/SEI-90-TR-021, 1990.

\bibitem{Thum:2014}
T.~Th\"{u}m, S.~Apel, C.~K\"{a}stner, I.~Schaefer, and G.~Saake, ``{A
  Classification and Survey of Analysis Strategies for Software Product
  Lines},'' \emph{ACM Computer Surveys}, vol.~47, no.~1, pp. 6:1--6:45, Jun.
  2014.

\bibitem{ShahinTSE}
R.~Shahin, M.~Akhundov, and M.~Chechik, ``{Annotative Software Product Line
  Analysis Using Variability-Aware Datalog},'' \emph{IEEE Transactions on
  Software Engineering}, vol.~49, no.~3, pp. 1323--1341, 2023.

\bibitem{datalog-defuse}
M.~Young and M.~Pezze, \emph{Software Testing and Analysis: Process, Principles
  and Techniques}.\hskip 1em plus 0.5em minus 0.4em\relax John Wiley \& Sons,
  Inc., 2005.

\bibitem{davis2010whence}
I.~J. Davis and M.~W. Godfrey, ``{From Whence It Came: Detecting Source Code
  Clones by Analyzing Assembler},'' in \emph{Proceedings of the Working
  Conference on Reverse Engineering (WCRE'10)}, 2010, pp. 242--246.

\bibitem{CPPX}
\BIBentryALTinterwordspacing
CPPX, ``{CPPX - C++ Fact Extractor},'' {Accessed: April 10, 2023}. [Online].
  Available: \url{https://www.swag.uwaterloo.ca/cppx/index.html}
\BIBentrySTDinterwordspacing

\bibitem{bravenboer2009strictly}
M.~Bravenboer and Y.~Smaragdakis, ``{Strictly Declarative Specification of
  Sophisticated Points-To Analyses},'' in \emph{Proceedings of the ACM SIGPLAN
  Conference on Object Oriented Programming Systems Languages and Applications
  (OOPSLA '09)}, 2009, pp. 243--262.

\bibitem{muscedere2019detecting}
B.~J. Muscedere, R.~Hackman, D.~Anbarnam, J.~M. Atlee, I.~J. Davis, and M.~W.
  Godfrey, ``{Detecting Feature-Interaction Symptoms In Automotive Software
  Using Lightweight Analysis},'' in \emph{Proceedings of the IEEE International
  Conference on Software Analysis, Evolution and Reengineering (SANER'19)},
  2019, pp. 175--185.

\bibitem{Neo4jBrowser}
``Neo4j browser,'' \url{https://neo4j.com/developer/neo4j-browser/}, accessed:
  2023-08-01.

\bibitem{grant1948latin}
D.~A. Grant, ``The latin square principle in the design and analysis of
  psychological experiments.'' \emph{Psychological bulletin}, vol.~45, no.~5,
  p. 427, 1948.

\bibitem{sauro2009comparison}
J.~Sauro and J.~S. Dumas, ``{Comparison of Three One-question, Post-task
  Usability Questionnaires},'' in \emph{Proceedings of the SIGCHI Conference on
  Human Factors in Computing Systems}, 2009, pp. 1599--1608.

\bibitem{allen2007likert}
I.~E. Allen and C.~A. Seaman, ``Likert scales and data analyses,''
  \emph{Quality progress}, vol.~40, no.~7, pp. 64--65, 2007.

\bibitem{leffingwell2000managing}
D.~Leffingwell and D.~Widrig, \emph{Managing Software Requirements: A Unified
  Approach}.\hskip 1em plus 0.5em minus 0.4em\relax Addison-Wesley
  Professional, 2000.

\bibitem{wilde2018merge}
E.~Wilde and D.~German, ``{Merge-Tree: Visualizing the Integration of Commits
  into Linux},'' \emph{Journal of Software: Evolution and Process}, vol.~30,
  no.~2, p.~25, 2018.

\bibitem{beck2011visual}
M.~Beck, J.~Tr{\"u}mper, and J.~D{\"o}llner, ``{A Visual Analysis and Design
  Tool for Planning Software Reengineerings},'' in \emph{Proceedings of the
  International Workshop on Visualizing Software for Understanding and Analysis
  (VISSOFT'11)}, 2011, pp. 1--8.

\bibitem{storey2001shrimp}
M.-A. Storey, C.~Best, and J.~Michand, ``{Shrimp Views: An Interactive
  Environment for Exploring Java Programs},'' in \emph{Proceedings of the
  International Workshop on Program Comprehension (IWPC'01)}, 2001, pp.
  111--112.

\bibitem{daniel2014polyptychon}
D.~T. Daniel, E.~Wuchner, K.~Sokolov, M.~Stal, and P.~Liggesmeyer,
  ``Polyptychon: A hierarchically-constrained classified dependencies
  visualization,'' in \emph{Proceedings of the IEEE Working Conference on
  Software Visualization (VISSOFT'14)}, 2014, pp. 83--86.

\bibitem{bennett2007working}
C.~Bennett, D.~Myers, M.-A. Storey, and D.~German, ``{Working with `Monster'
  Traces: Building a Scalable, Usable, Sequence Viewer},'' in \emph{In
  Proceedings of the International Workshop on Program Comprehension Through
  Dynamic Analysis (PCODA'07)}, 2007, pp. 1--5.

\bibitem{borowski2022graph}
K.~Borowski, B.~Balis, and T.~Orzechowski, ``{Graph Buddy—an Interactive Code
  Dependency Browsing and Visualization Tool},'' in \emph{Proceedings of the
  Working Conference on Software Visualization (VISSOFT'22)}, 2022, pp.
  152--156.

\bibitem{latoza2011visualizing}
T.~D. LaToza and B.~A. Myers, ``{Visualizing Call Graphs},'' in
  \emph{Proceedings of the IEEE Symposium on Visual Languages and Human-Centric
  Computing (VL/HCC'11)}, 2011, pp. 117--124.

\bibitem{arora2019javarelationshipgraphs}
R.~Arora and S.~Goel, ``{JavaRelationshipGraphs (JRG) Transforming Java
  Projects into Graphs using Neo4j Graph Databases},'' in \emph{Proceedings of
  the International Conference on Software Engineering and Information
  Management (ICSIM'19)}, 2019, pp. 80--84.

\bibitem{kastner2009guaranteeing}
C.~K{\"a}stner, S.~Apel, S.~Trujillo, M.~Kuhlemann, and D.~Batory,
  ``{Guaranteeing Syntactic Correctness for All Product Line Variants: A
  Language-Independent Approach},'' in \emph{Proceedings of the Interntional
  Conferfence on Objects, Components, Models and Patterns (TOOLS'09))}, 2009,
  pp. 175--194.

\bibitem{Heidenreich-VISPLE08}
F.~Heidenreich, I.~\c{S}avga, and C.~Wende, ``{On Controlled Visualisations in
  Software Product Line Engineering},'' in \emph{Proceedings of the
  International Workshop on Visualisation of Software Product Line Engineering
  (ViSPLE@SPLC'08)}, 2008, pp. 335--341.

\bibitem{czarnecki2006verifying}
K.~Czarnecki and K.~Pietroszek, ``{Verifying Feature-Based Model Templates
  against Well-Formedness OCL Constraints},'' in \emph{Proceedings of
  Generative Programming: Concepts and Experiences (GPCE'06)}, 2006, pp.
  211--220.

\bibitem{Apel:2011}
S.~Apel and D.~Beyer, ``{Feature Cohesion in Software Product Lines: An
  Exploratory Study},'' in \emph{Proceedings of the ACM/IEEE International
  Conference on Software Engineering (ICSE'11)}.\hskip 1em plus 0.5em minus
  0.4em\relax New York, NY, USA: ACM, 2011, pp. 421--430.

\bibitem{asadi2016effects}
M.~Asadi, S.~Soltani, D.~Ga{\v{s}}evi{\'c}, and M.~Hatala, ``{The Effects of
  Visualization and Interaction Techniques on Feature Model Configuration},''
  \emph{Empirical Software Engineering}, vol.~21, no.~4, pp. 1706--1743, 2016.

\bibitem{loesch2007optimization}
F.~Loesch and E.~Ploedereder, ``{Optimization of Variability in Software
  Product Lines},'' in \emph{Proceedings of International Conference on
  Software Product Lines (SPLC'07)}, 2007, pp. 151--162.

\bibitem{botterweck2008visual}
G.~Botterweck, S.~Thiel, D.~Nestor, S.~bin Abid, and C.~Cawley, ``{Visual Tool
  Support for Configuring and Understanding Software Product Lines},'' in
  \emph{Proceedings of the International Conference on Software Product Lines
  (SPLC'08}, 2008, pp. 77--86.

\bibitem{martinez2014feature}
J.~Martinez, T.~Ziadi, R.~Mazo, T.~F. Bissyand{\'e}, J.~Klein, and Y.~Le~Traon,
  ``{Feature Relations Graphs: A Visualisation Paradigm for Feature Constraints
  in Software Product Lines},'' in \emph{Proceedings of the IEEE Working
  Conference on Software Visualization (VISSOFT'14)}.\hskip 1em plus 0.5em
  minus 0.4em\relax IEEE, 2014, pp. 50--59.

\bibitem{struber2020variability}
D.~Str{\"u}ber, A.~Anjorin, and T.~Berger, ``{Variability Representations in
  Class Models: An Empirical Assessment},'' in \emph{Proceedings of the
  ACM/IEEE International Conference on Model Driven Engineering Languages and
  Systems (MODELS'20)}, 2020, pp. 240--251.

\end{thebibliography}

\end{document}